\documentclass[11pt,aps,prd,nofootinbib,superscriptaddress,preprint,preprintnumbers]{revtex4-2}

\usepackage{amsmath}
\usepackage{amssymb}
\usepackage{graphicx}
\usepackage{xcolor}
\usepackage{hyperref}
\usepackage{bbm}
\usepackage{makecell}
\usepackage{multirow}
\usepackage{tikz}
\usetikzlibrary{backgrounds}
\usepackage{float}

\allowdisplaybreaks

\newcommand{\SKLP}{State Key Laboratory of Particle Detection and Electronics, University of Science and Technology of China, Hefei 230026, Anhui, People’s Republic of China}
\newcommand{\USTC}{Department of Modern Physics, University of Science and Technology of China, Hefei 230026, Anhui, People’s Republic of China}
\newcommand{\ACFS}{Anhui Center for Fundamental Sciences in Theoretical Physics, University of Science and Technology of China, Hefei 230026, Anhui, People’s Republic of China}
\newcommand{\NUS}{Department of Physics, National University of Singapore, Singapore 117551, Singapore}

\newcommand{\Hline}{\noalign{\hrule height 1.2pt}}

\begin{document}

\title{\boldmath Planar master integrals for two-loop NLO electroweak light-fermion contributions to $g g \rightarrow Z H$}

\author{Shu-Xiang Li}
\affiliation{\SKLP}
\affiliation{\USTC}

\author{Ren-You Zhang}
\email{zhangry@ustc.edu.cn}
\affiliation{\SKLP}
\affiliation{\USTC}
\affiliation{\ACFS}

\author{Xiao-Feng Wang}
\affiliation{\SKLP}
\affiliation{\USTC}

\author{Pan-Feng Li}
\affiliation{\SKLP}
\affiliation{\USTC}

\author{Xiang-Jie Wei}
\affiliation{\SKLP}
\affiliation{\USTC}

\author{Yi Wang}
\affiliation{\SKLP}
\affiliation{\USTC}

\author{Yi Jiang}
\affiliation{\SKLP}
\affiliation{\USTC}

\author{Qing-hai Wang}
\email{qhwang@nus.edu.sg}
\affiliation{\NUS}

\date{\today}
\begin{abstract}
For the two-loop next-to-leading-order electroweak (NLO EW) corrections to $gg \rightarrow ZH$, the light-fermion contributions can be classified into eight distinct topologies. Using the canonical differential-equations method, we perform an analytic computation of the master integrals (MIs) associated with the four planar topologies. Canonical bases are constructed using the Magnus-expansion method, and the resulting alphabets consist of algebraic symbol letters involving nontrivial radicals. We develop a systematic framework for identifying the radical structures of the canonical MIs, enabling their organization into suitable subsystems and, whenever possible, their representation in terms of Goncharov polylogarithms (GPLs) up to $\mathcal{O}(\epsilon^4)$. Only a few MIs at $\mathcal{O}(\epsilon^3)$ and $\mathcal{O}(\epsilon^4)$ are instead represented as one-fold integrals over GPLs, due to the presence of nested square roots that obstruct the simultaneous rationalization of all radicals.

\begin{description}
\item[keywords]
Master integrals, Canonical differential-equations method, $ZH$ associated production
\end{description}
\end{abstract}

\maketitle
\newpage

\section{Introduction}
The observation of a Higgs boson with a mass of $125~ \mathrm{GeV}$ at the Large Hadron Collider (LHC) in 2012 completed the particle spectrum of the standard model (SM), marking a major triumph of the theory. At the LHC, the dominant Higgs boson production mechanisms are gluon-gluon fusion ($gg\text{F}$), vector-boson fusion (VBF), associated production with a weak gauge boson ($VH$), and associated production with a top-quark pair ($t\bar{t}H$). Among these, the $VH$ production mode plays a central role in the experimental observation of the $H \rightarrow b\bar{b}$ decay at ATLAS \cite{ATLAS:2018kot} and CMS \cite{CMS:2018nsn}. This is primarily due to the fact that the accompanying $W$ or $Z$ boson, when decaying leptonically, provides clean and efficient trigger signatures, while substantially reducing the otherwise overwhelming multijet background. Moreover, the $pp \rightarrow VH$ process is sensitive to the strength of the $VVH$ coupling and is particularly well suited for probing potential deviations from SM predictions \cite{CMS:2021nnc,Rao:2023ogi}. As such, it constitutes a key channel for testing the gauge structure of the SM.

\par
At hadron colliders, the leading-order contribution to $VH$ associated production arises from Drell-Yan-type quark-antiquark annihilation. The QCD corrections to $VH$ production have been computed up to next-to-next-to-next-to-leading order ($\text{N}^3\text{LO}$) at the inclusive level \cite{Baglio:2022wzu}, with next-to-next-to-leading order (NNLO) predictions available through the \texttt{VH@NNLO} package \cite{Brein:2003wg, Brein:2011vx, Brein:2012ne}, which is widely employed in precision phenomenology. The next-to-leading order (NLO) electroweak (EW) corrections have been calculated in Refs.\cite{Ciccolini:2003jy, Denner:2011id} and implemented in the \texttt{HAWK} Monte Carlo program \cite{Denner:2014cla}. Combined NLO QCD and EW corrections are presented in Refs.\cite{Granata:2017iod, Obul:2018psx}. The NNLO QCD predictions for the differential distributions of $WH$ and $ZH$ production are available in Refs.\cite{Ferrera:2011bk, Ferrera:2014lca}. These fixed-order results have been matched to parton showers to achieve NNLO+PS accuracy in exclusive event simulations \cite{Astill:2016hpa,Astill:2018ivh}, allowing for precise comparisons with experimental data. In the five-flavor scheme, the $b\bar{b} \rightarrow ZH$ process, incorporating nonzero bottom-quark Yukawa coupling, has been computed at QCD NNLO within the soft-virtual approximation \cite{Ahmed:2019udm}. Moreover, the polarized $q \bar{q} \rightarrow ZH$ amplitudes at two-loop order in QCD have been studied analytically in Ref.\cite{Ahmed:2020kme}.

\par
The loop-induced process $gg \rightarrow ZH$ contributes at leading order with a cross section proportional to $\alpha_s^2$, thereby constituting a NNLO QCD correction to the $pp \rightarrow ZH$ process. However, owing to the large gluon luminosity at the LHC, this loop-induced channel contributes appreciably, representing approximately $10\%$ of the total $ZH$ cross section \cite{ATLAS:2016neq}. In the boosted regime, where the Higgs boson possesses a large transverse momentum, this relative contribution is further enhanced \cite{Harlander:2013mla,Englert:2013vua}, making it a notable source of theoretical uncertainty. Moreover, since the amplitude depends on the particle content and couplings in the loops, this channel provides a particularly promising probe of potential new physics. The LO results for $ZH$ production via gluon-gluon fusion were first presented in Refs.\cite{Dicus:1988yh,Kniehl:1990iva}. The complete NLO QCD corrections were initially computed in the heavy-top limit ($m_t \rightarrow +\infty$) \cite{Altenkamp:2012sx}, resulting in a substantial enhancement of the cross section with a $K$-factor close to $2$, and the finite top-quark mass effects have subsequently been analyzed using large-$m_t$ expansions \cite{Hasselhuhn:2016rqt}. Two-loop virtual corrections computed via a transverse-momentum expansion are presented in Ref.\cite{Alasfar:2021ppe}, while results in the high-energy and large-$m_t$ limits have been reported in Ref.\cite{Davies:2020drs}. Fully mass-dependent two-loop amplitudes have also been evaluated numerically \cite{Chen:2020gae}. Building on these results, full NLO QCD predictions have been further refined through a variety of expansion techniques, including small-$m_H$ and $m_Z$ expansions \cite{Wang:2021rxu}, combinations of sector decomposition with high-energy expansions \cite{Chen:2022rua}, and hybrid approaches that integrate small-$p_T$ and high-energy expansions to achieve comprehensive coverage of the phase space \cite{Degrassi:2022mro,Bellafronte:2022jmo}.

\par
Although QCD corrections to the gluon-fusion channel in $ZH$ production have been extensively studied, a comprehensive analysis of the corresponding NLO EW corrections remains lacking. These corrections involve virtual contributions with a large number of diagrams that introduce an additional mass scale, $m_W$, rendering their computation highly nontrivial. For reference, the NLO EW corrections to $gg \rightarrow gH$ computed in Ref.\cite{Bi:2025oga} suggest that, in processes analogous to $gg \rightarrow ZH$, EW effects at NLO contribute approximately $4\%$ to the total cross section. Therefore, such contributions are expected to be non-negligible in future high-precision predictions for $gg \rightarrow ZH$.

\par
This study presents an analytic evaluation of the master integrals (MIs) for the NLO EW corrections to $gg \rightarrow ZH$, focusing on the planar topologies with light-fermion loops. The calculation is performed within the framework of canonical differential equations, yielding analytic $\epsilon$-expansions of these integrals up to $\mathcal{O}(\epsilon^4)$. The coefficients are expressed in terms of Goncharov polylogarithms (GPLs) wherever feasible, and otherwise as one-fold integrals over GPLs. The rest of this paper is organized as follows. In Section \ref{sec:2}, we present the kinematics of $gg \rightarrow ZH$ and introduce the topologies of the two-loop integrals involved. The differential-equations method adopted in our calculation is detailed in Section \ref{sec:3}. Sections \ref{sec:4} and \ref{sec:5} are devoted to the analytic evaluation of the canonical bases for two family branches, respectively. In Section \ref{sec:6}, we validate our analytic expressions by comparing them with numerical results. Finally, a brief summary is provided in Section \ref{sec:7}.

\section{Kinematics and topologies}
\label{sec:2}
\par
In this work, we focus on the two-loop NLO EW virtual corrections to $ZH$ production via gluon-gluon fusion,
\begin{equation}
g(p_{1}) + g(p_{2}) \rightarrow Z(p_{3}) + H(p_{4})\,,
\end{equation}
where all four external momenta $p_i$ ($i = 1, \ldots, 4$) are defined as incoming. The amplitude for this $2 \rightarrow 2$ scattering process can be conveniently expressed in terms of the Mandelstam invariants,
\begin{equation}
s = {(p_{1} + p_{2})}^{2}\,,
\qquad
t = {(p_{2} + p_{3})}^{2}\,,
\qquad
u = {(p_{1} + p_{3})}^{2}\,.
\end{equation}
Momentum conservation and the on-shell conditions for the external momenta imply $s + t + u = m_Z^2 + m_H^2$. Using this relation, we adopt either $\{s,\, t\}$ or $\{u,\, t\}$ as the independent Mandelstam variables in the subsequent analysis, depending on the topology under consideration.

\par
\begin{figure}[htbp]
  \centering
  \begin{minipage}{0.23\textwidth}
    \includegraphics[width=1.0\textwidth]{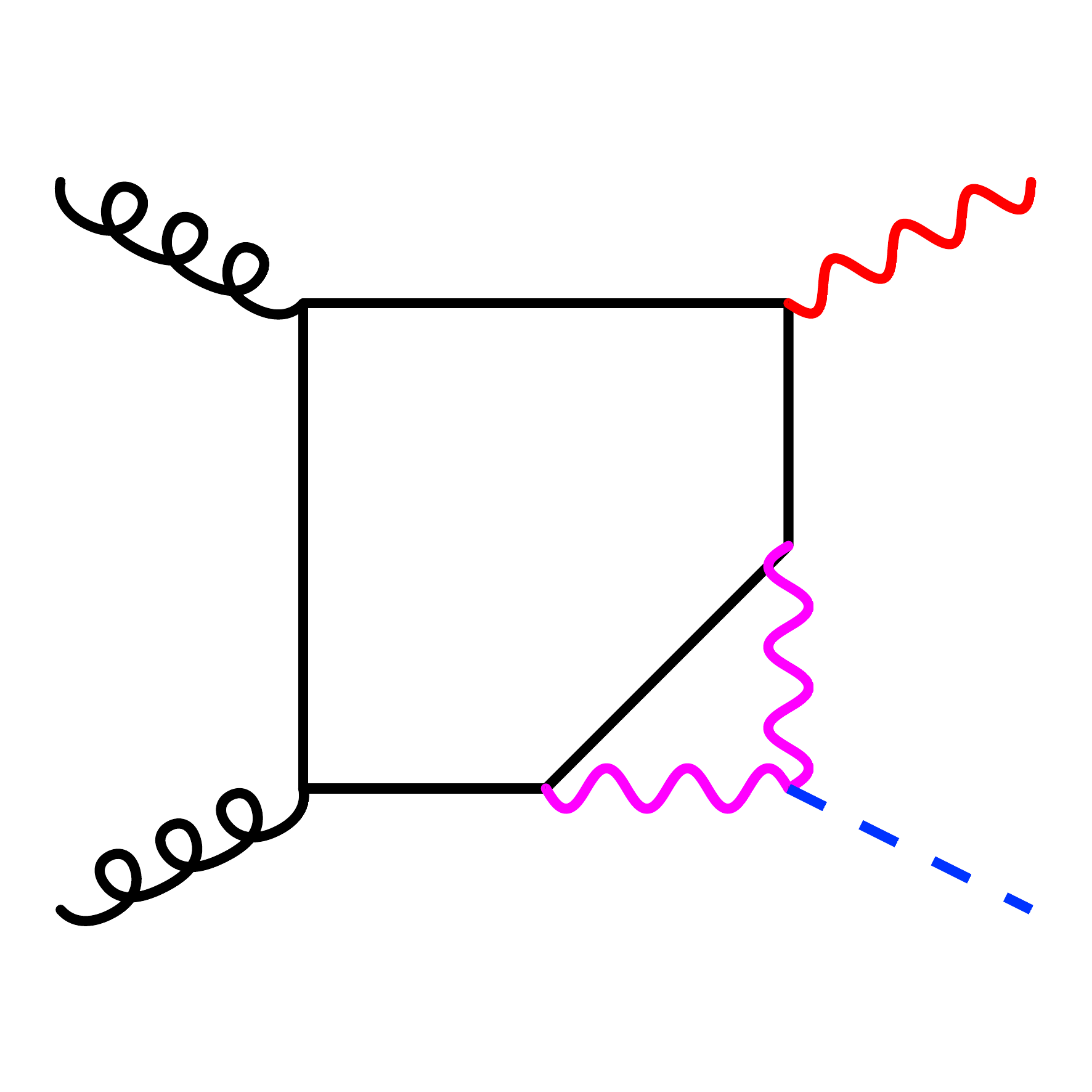}
  \end{minipage}
  \begin{minipage}{0.23\textwidth}
    \includegraphics[width=1.0\textwidth]{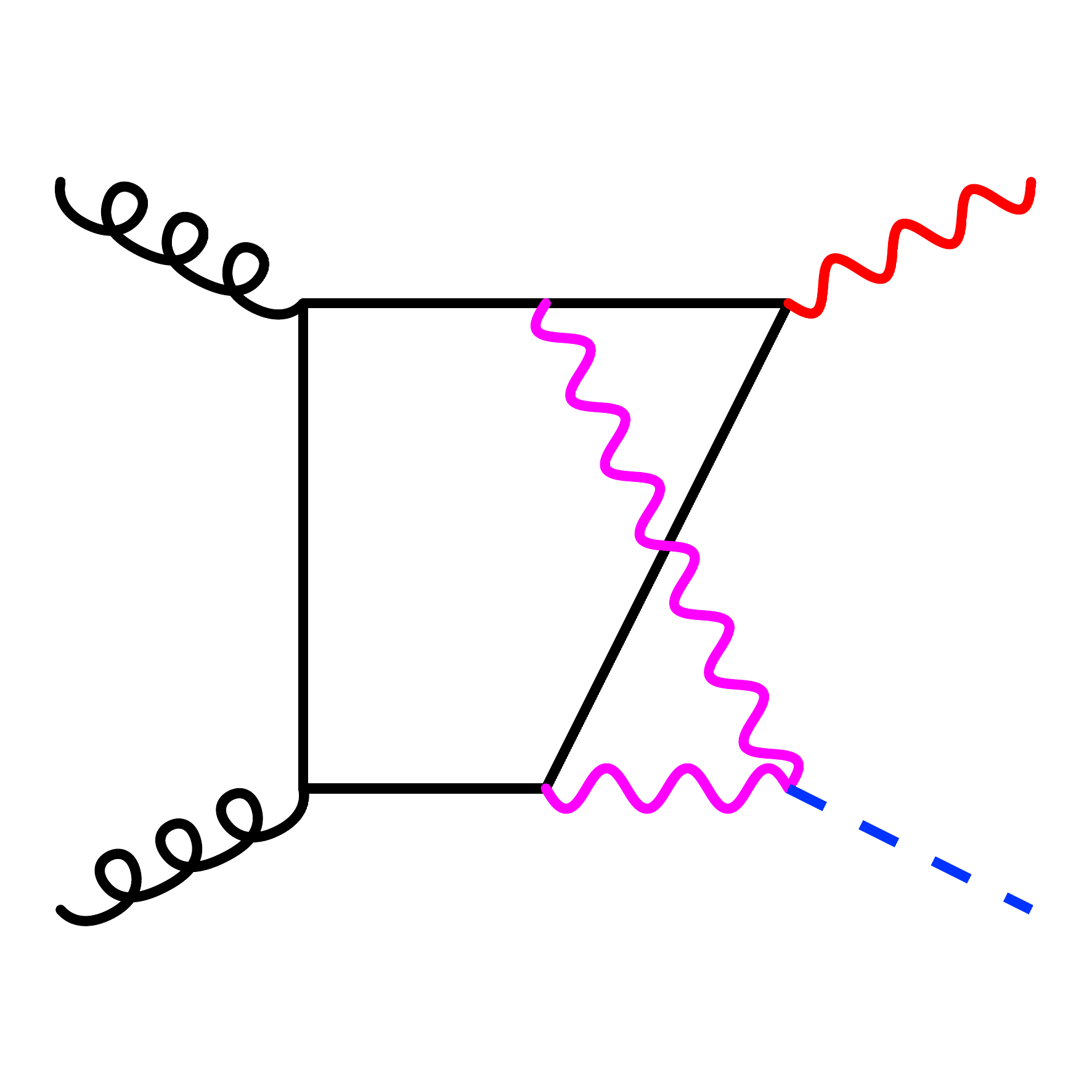}
  \end{minipage}
  \begin{minipage}{0.23\textwidth}
    \includegraphics[width=1.0\textwidth]{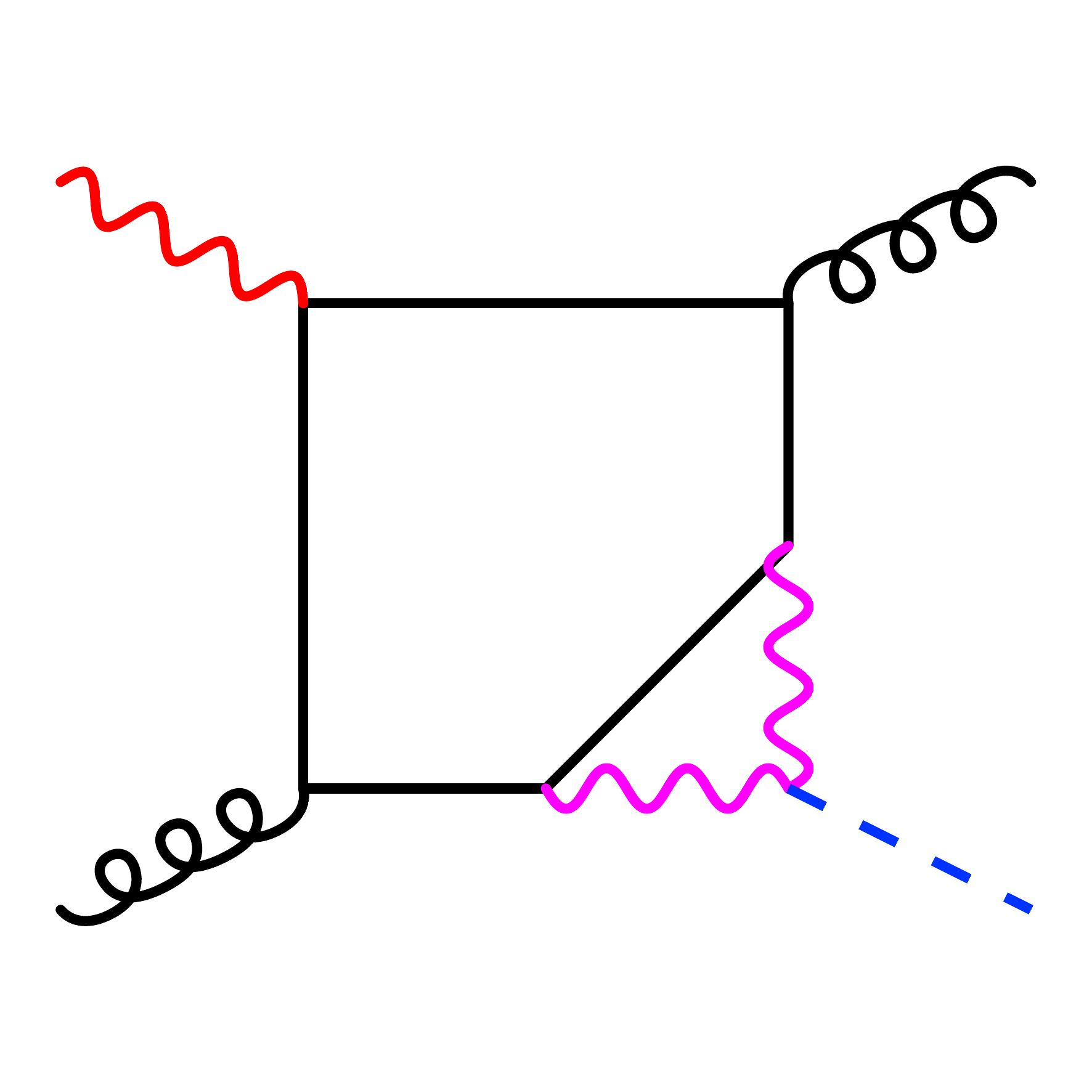}
  \end{minipage}
  \begin{minipage}{0.23\textwidth}
    \includegraphics[width=1.0\textwidth]{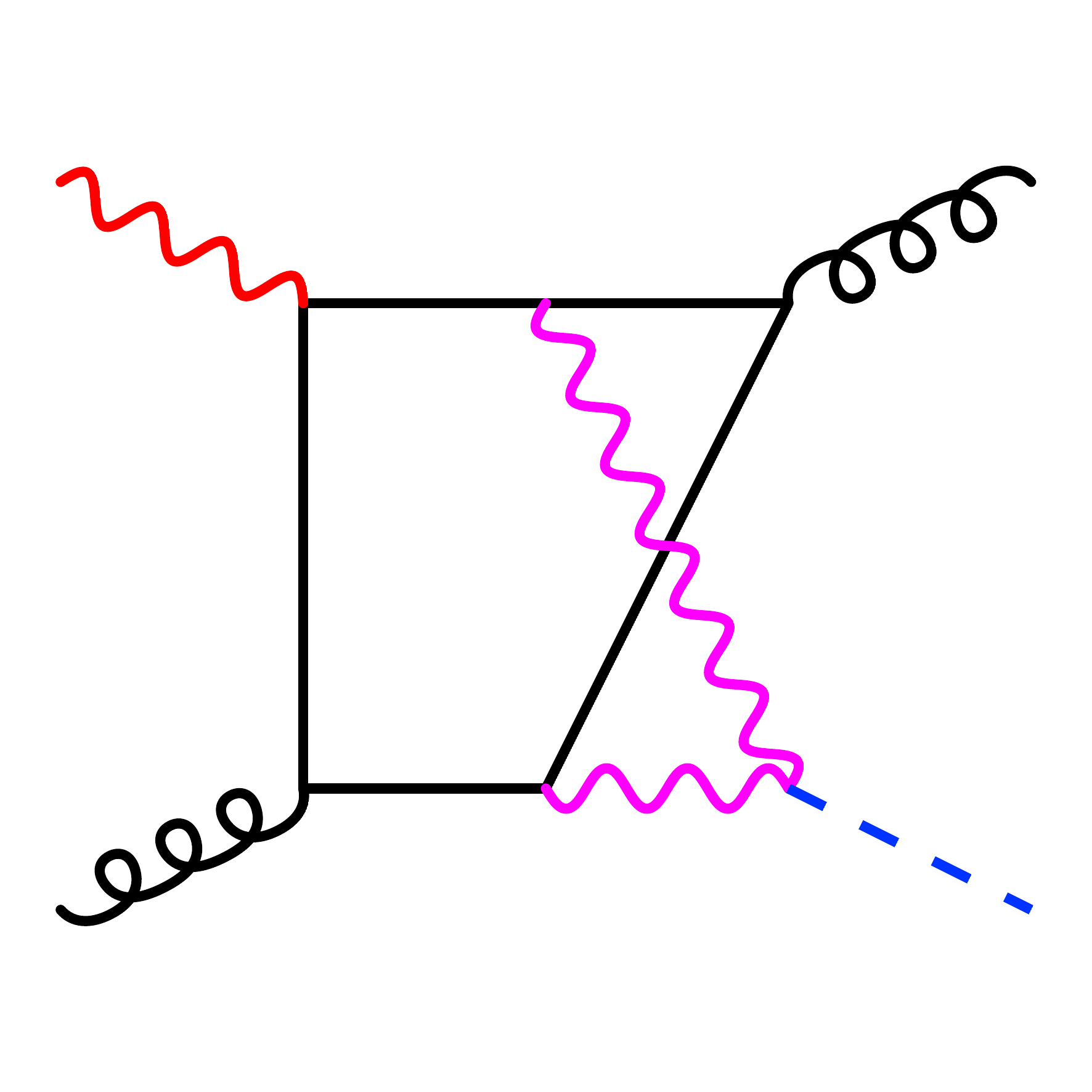}
  \end{minipage}
\caption{
Four Feynman-diagram topologies contributing to the NLO EW virtual corrections to $gg \rightarrow ZH$, induced by light quarks and involving the $WWH$ vertex. Line styles denote particle species: solid black for light fermions, curly black for gluons, wavy pink for $W$ bosons, wavy red for $Z$ bosons, and dashed blue for Higgs bosons.
}
\label{fig1}
\end{figure}

For the $gg \rightarrow ZH$ process, a total of $132$ irreducible two-loop Feynman diagrams induced by light quarks contribute to the NLO EW corrections. These diagrams can be systematically classified into eight distinct topologies: four involving the $WWH$ vertex and four analogous ones in which the $WWH$ vertex is replaced by $ZZH$. Figure \ref{fig1} illustrates the first four topologies, comprising two planar and two non-planar configurations, with only the planar diagrams considered in the present calculation.

\par
Within dimensional regularization, the dimensionless two-loop four-point scalar Feynman integrals belonging to the family $\mathcal{F}(D_1, \dots, D_9)$ are conventionally defined as
\begin{equation}
I(a_{1}, \ldots, a_{9})
=
\frac{e^{2 \epsilon \gamma_{E}}}{(Q^2)^{d - a}}
\int \frac{d^{d} l_{1}}{i \pi^{d / 2}}\, \frac{d^{d} l_{2}}{i \pi^{d / 2}}\, \frac{1}{{D}_1^{a_1} \cdots D_{9}^{a_9}}\,,
\end{equation}
where $\gamma_E$ denotes the Euler-Mascheroni constant, $d = 4 - 2\epsilon$ specifies the spacetime dimension, $a = a_1 + \cdots + a_9$, and $Q$ is a characteristic scale introduced to render $I$ dimensionless. The loop momenta are labeled $l_1$ and $l_2$, and $D_i$ ($i = 1, \dots, 9$) denote the propagators of the internal lines. Given a set of propagators $\{D_i\}$, all integrals with arbitrary indices $\{a_i\}$ constitute an integral family. According to the signs of the propagator indices $\{a_i\}$, an integral family can be decomposed into sectors. To this end, we introduce a binary vector $(s_1, \ldots, s_9)$, with $s_i \in \{0,\, 1\}$, to label the sectors of the integral family $\mathcal{F}$. The sector $\mathcal{S}_{\mathcal{F}}(s_1, \ldots, s_9)$ is defined as
\begin{equation}
\mathcal{S}_{\mathcal{F}}(s_1, \ldots, s_9)
=
\Big\{
I(a_1, \ldots, a_9) \mid \Theta(a_i) = s_i,\, i = 1, \dots, 9
\Big\}\,,
\end{equation}
where $\Theta(x)$ denotes the Heaviside step function, defined by $\Theta(x)=1$ for $x>0$ and $\Theta(x)=0$ for $x\leqslant 0$. A sector $\mathcal{S}_{\mathcal{F}}(s_1, \ldots, s_9)$ is called a sub-sector of $\mathcal{S}_{\mathcal{F}}(t_1, \ldots, t_9)$ if $s_i \leqslant t_i$ for all $i = 1, \ldots, 9$. The union of $\mathcal{S}_{\mathcal{F}}(s_1, \ldots, s_9)$ and all its sub-sectors defines the branch induced by $\mathcal{S}_{\mathcal{F}}(s_1, \ldots, s_9)$, denoted by $\mathcal{B}_{\mathcal{F}}(s_1, \ldots, s_9)$. The four topologies associated with the $WWH$ vertex shown in Fig.\ref{fig1} fall into two distinct integral families:
\begin{enumerate}
\item Family $\mathcal{F}_1$:
\begin{align}
&
D_{1} = l_{1}^{2}
&&
D_{2} = (l_{1} + p_{1})^{2}
&&
D_{3} = (l_{1} - p_{3} - p_{4})^{2}
\nonumber
\\
&
D_{4} = (l_{1} - p_{4})^{2}
&&
D_{5} = (l_{2} - p_{4})^{2} - m_{W}^{2}
&&
D_{6} = l_{2}^{2} - m_{W}^{2}
\\
&
D_{7} = (l_{1} - l_{2})^{2}
&&
D_{8} = (l_{1} - l_{2} - p_{3})^{2}
&&
D_{9} = (l_{2} + p_{1})^{2}
\nonumber
\end{align}
two topologies $\subset \mathcal{F}_1$:
\begin{equation}
\mathcal{S}_1 \equiv \mathcal{S}_{\mathcal{F}_1}(1, 1, 1, 1, 1, 1, 1, 0, 0)\,,
\qquad
\mathcal{S}_{1N} \equiv \mathcal{S}_{\mathcal{F}_1}(1, 1, 1, 0, 1, 1, 1, 1, 0)
\end{equation}
\item Family $\mathcal{F}_2$:
\begin{align}
&
D_{1} = l_{1}^{2}
&&
D_{2} = (l_{1} + p_{1})^{2}
&&
D_{3} = (l_{1} - p_{2} - p_{4})^{2}
\nonumber
\\
&
D_{4} = (l_{1} - p_{4})^{2}
&&
D_{5} = (l_{2} - p_{4})^{2} - m_{W}^{2}
&&
D_{6} = l_{2}^{2} - m_{W}^{2}
\\
&
D_{7} = (l_{1} - l_{2})^{2}
&&
D_{8} = (l_{1} - l_{2} + p_{1})^{2}
&&
D_{9} = (l_{2} + p_{2})^{2}
\nonumber
\end{align}
two topologies $\subset \mathcal{F}_2$:
\begin{equation}
\mathcal{S}_{2} \equiv \mathcal{S}_{\mathcal{F}_2}(1, 1, 1, 1, 1, 1, 1, 0, 0)\,,
\qquad
\mathcal{S}_{2N} \equiv \mathcal{S}_{\mathcal{F}_2}(0, 1, 1, 1, 1, 1, 1, 1, 0)
\end{equation}
\end{enumerate}
Here, $\mathcal{S}_1$ and $\mathcal{S}_2$ correspond to the planar topologies shown in Fig.\ref{fig1}, which are evaluated analytically in this work, whereas $\mathcal{S}_{1N}$ and $\mathcal{S}_{2N}$ denote the remaining two non-planar topologies.

\par
Integration-by-parts (IBP) \cite{Tkachov:1981wb,Chetyrkin:1981qh} identities establish linear relations among integrals within a given branch of an integral family, with coefficients given by rational functions of the external kinematic invariants and internal masses. These relations allow all integrals within a given branch to be reduced to a finite set of master integrals (MIs) \cite{Smirnov:2010hn}. For convenience, we denote the branch induced by $\mathcal{S}_1$ as $\mathcal{B}_1$, and that by $\mathcal{S}_2$ as $\mathcal{B}_2$. Using the \texttt{Kira} package \cite{Maierhofer:2017gsa,Klappert:2020nbg}, which implements the Laporta algorithm \cite{Laporta:2000dsw}, we have reduced the integrals in $\mathcal{B}_{1}$ and $\mathcal{B}_{2}$ to $62$ and $59$ MIs, respectively.

\section{Differential-equations method}
\label{sec:3}
\par
The derivative of any scalar Feynman integral with respect to kinematic invariants can be reduced, via IBP identities, to linear combinations of integrals within the same family branch, implying that the branch is closed under differentiation with respect to these invariants. Consequently, the corresponding MIs satisfy a closed system of linear differential equations in these variables. We employ \texttt{LiteRed} \cite{Lee:2012cn,Lee:2013mka} to perform the differentiation, and use \texttt{Kira} to carry out the reduction, thereby obtaining a coupled system of linear differential equations satisfied by the MIs. However, obtaining analytic solutions to such multivariable systems of differential equations is, in general, a highly nontrivial task. If there exists a basis $\vec{g}(\vec{x}, \epsilon)$ in which the differential equations can be cast into a canonical (or $\epsilon$-factorized) form,
\begin{equation}
\label{eq:canonical-DEs}
d \vec{g}(\vec{x}, \epsilon) = \epsilon\, d \mathbbm{A}(\vec{x})\, \vec{g}(\vec{x}, \epsilon)\,,
\end{equation}
the system simplifies considerably, and its solution can be written in terms of Chen's iterated integrals \cite{Chen:1977oja},
\begin{equation}
\label{eq:Chen-int}
\vec{g}(\vec{x}, \epsilon) = \mathcal{P} \exp\Big(\epsilon \int_{\gamma} d \mathbbm{A}\Big)\, \vec{g}(\vec{x}_0, \epsilon)\,,
\end{equation}
where $\mathcal{P}$ denotes the path-ordering operator, and $\gamma$ is an integration path connecting $\vec{x}_0$ to $\vec{x}$. This approach is known as the canonical differential-equations method \cite{Henn:2013pwa}, and the corresponding set of MIs is referred to as a canonical basis. Several approaches have been developed for constructing a canonical basis, including the Magnus expansion \cite{Magnus:1954zz,Blanes:2008xlr,Argeri:2014qva}, $d\log$ integrand construction \cite{Henn:2020lye}, intersection theory \cite{Mastrolia:2018uzb,Chen:2020uyk}, and leading singularity analysis  \cite{Henn:2013pwa,Henn:2014qga}. In this work, the canonical basis is constructed via the Magnus expansion starting from a linear basis.

\par
Let $\vec{f}(\vec{x}, \epsilon)$ be a linear basis of a family branch of Feynman integrals, satisfying the following system of differential equations:
\begin{equation}
\label{eq:linear-DEs}
\partial_i \vec f(\vec{x}, \epsilon)
=
\Big[\,
\mathbbm{B}_i^{(0)}(\vec{x}) + \epsilon\, \mathbbm{B}_i^{(1)}(\vec{x})
\,\Big]\,
\vec{f}(\vec{x}, \epsilon)\,,
\qquad
(i = 1, \ldots, n)\,,
\end{equation}
where $\partial_i \equiv \partial/\partial x_i$ and $\vec{x} = (x_1, \ldots, x_n)$. We introduce the matrix $\mathbbm{T}$, which is defined recursively as follows:
\begin{equation}
\begin{aligned}
&
\mathbbm{T} = \mathbbm{T}_n\,,
\\
&
\end{aligned}
\quad~
\left\lbrace
\begin{aligned}
&
\mathbbm{T}_i = \mathbbm{T}_{i-1} \exp\Big( \Omega_i \big[ \widetilde{\mathbbm{B}}_i^{(0)} \big] \Big)
~~\text{with}~~
\widetilde{\mathbbm{B}}_i^{(0)} = \mathbbm{T}_{i-1}^{-1}\, \mathbbm{B}_i^{(0)}\, \mathbbm{T}_{i-1}
\quad
(i = 1, \ldots, n)
\\
&
\mathbbm{T}_0 = \mathbbm{1}_{n \times n}
\end{aligned}
\right.
\end{equation}
where $\Omega_i$ denotes a family of matrix-valued functionals indexed by $i \in \{1,\dots,n\}$, each admitting a series representation known as the Magnus expansion \cite{Argeri:2014qva},
\begin{align}
\Omega_i\big[ \mathbbm{F} \big](\vec{x})
=&
\int^{x_i} dx_i^{\prime}\, \mathbbm{F}(\cdots,\,x_{i-1},\, x_i^{\prime},\, x_{i+1},\, \cdots)
\\
&+
\frac{1}{2!}
\int^{x_i} dx_i^{\prime} \int^{x_i^{\prime}} dx_i^{\prime\prime}\,
\big[\, \mathbbm{F}(\cdots,\,x_{i-1},\, x_i^{\prime},\, x_{i+1},\, \cdots)\,,\,
\mathbbm{F}(\cdots,\,x_{i-1},\, x_i^{\prime\prime},\, x_{i+1},\, \cdots) \,\big]
\nonumber
\\
&+ \cdots
\nonumber
\end{align}
The matrix $\mathbbm{T}$ is invertible, and the transformation $\mathbbm{T}^{-1}$ maps the linear basis $\vec{f}$ to the canonical basis $\vec{g} = \mathbbm{T}^{-1}\vec{f}$. The validity of the recursive construction of this transformation is ensured by the integrability conditions of the linear differential system \eqref{eq:linear-DEs}. Accordingly, the canonical basis $\vec{g}$ satisfies the total differential equation \eqref{eq:canonical-DEs}, with $d\mathbbm{A}$ given by
\begin{equation}
\label{eq:dA-DEs}
d\mathbbm{A} = \sum_{i=1}^{n} \mathbbm{T}^{-1} \mathbbm{B}_i^{(1)} \mathbbm{T}\, dx_i\,.
\end{equation}
The differential one-form $d\mathbbm{A}$ can be written in the $d\log$ form,
\begin{equation}
d \mathbbm{A}(\vec{x}) = \sum_a\, \mathbbm{M}_a\, d \log{\omega_a(\vec{x})}\,,
\end{equation}
where $\mathbbm{M}_a$ are constant matrices, and $\omega_a$, referred to as symbol letters, are algebraic functions of $\vec{x}$ and collectively constitute the alphabet of the canonical differential system. Therefore, the recursion relations governing the coefficients in the $\epsilon$-expansion of the path-ordered integral representation of $\vec{g}(\vec{x}, \epsilon)$ take the form
\begin{equation}
\label{eq:recursion-relation}
\vec g^{\,(p)}(\vec{x}) = \vec g^{\,(p)}(\vec{x}_0)
+
\int_0^1 \sum_a \mathbbm{M}_a\, \frac{d \log \omega_a(\vec{x}(t))}{dt} \, \vec{g}^{\,(p-1)}(\vec{x}(t))\, dt\,,
\qquad
p = 1, 2, \ldots
\end{equation}
where $\vec{x}(t)$ parametrizes an integration path from $\vec{x}_0$ to $\vec{x}$, satisfying $\vec{x}(0) = \vec{x}_0$ and $\vec{x}(1) = \vec{x}$. If the alphabet consists of rational letters, the $d\log$ kernels in the integral representation of $\vec{g}^{\,(p)}$ reduce to rational functions with simple poles. The corresponding iterated integrals can therefore be expressed in terms of GPLs, which are defined recursively by
\begin{equation}
G(a_n, \ldots, a_1; z) = \int_0^z \frac{dz^{\prime}}{z^{\prime} - a_n}\, G(a_{n-1}, \ldots, a_1; z^{\prime})\,,
\end{equation}
with $G(\, ; z) = 1$ and $G(\vec{0}_n; z) = \log^n (z) / n!$.

\section{Family branch $\mathcal{B}_{1}$}
\label{sec:4}
\par
\begin{figure}[htbp]
\centering
\begin{minipage}{0.95\textwidth}
\includegraphics[width=1.0\textwidth]{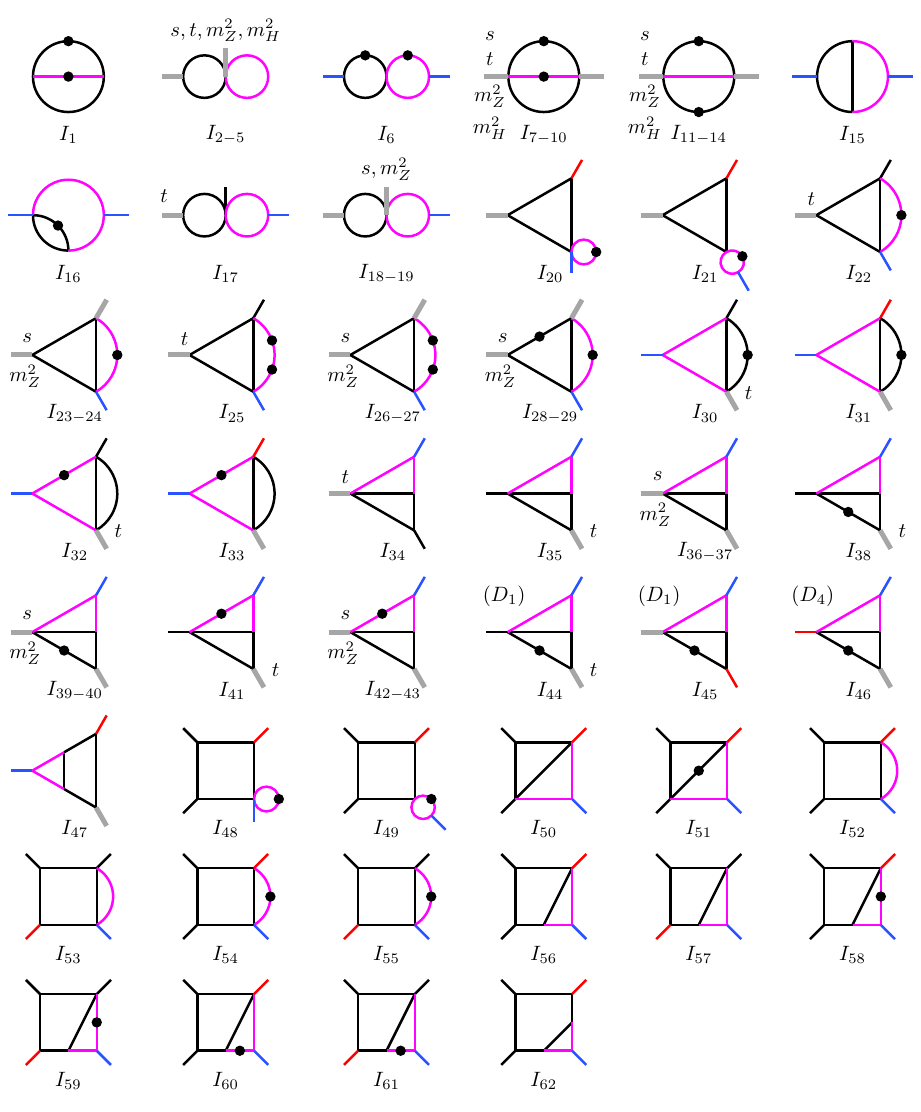}
\end{minipage}
\caption{
Pre-canonical basis for the $\mathcal{B}_1$ branch. Line colors follow the same convention as in Fig.\ref{fig1}. Each dot on a propagator denotes an increment of one in the power of that propagator. The label $(D_i)$ in the upper-left corner indicates that propagator $D_i$ carries power $-1$, while the labels adjacent to the thick gray external lines specify the corresponding kinematic invariants.
}
\label{fig2}
\end{figure}

The Feynman integrals in the $\mathcal{B}_{1}$ branch depend on five independent kinematic invariants with mass-squared dimension, which we take to be $\big\{ s,\, t,\, m_Z^2,\, m_H^2,\, m_W^2 \big\}$. Upon fixing the characteristic scale to $Q = m_W$, the dimensionless integrals can be conveniently parametrized in terms of the following dimensionless kinematic variables:
\begin{equation}
x = -\, \frac{s}{m_W^2}\,,
\qquad
y = -\, \frac{t}{m_W^2}\,,
\qquad
z = -\, \frac{m_Z^2}{m_W^2}\,,
\qquad
w = -\, \frac{m_H^2}{m_W^2}\,,
\end{equation}
which we collectively denote by the vector $\vec{x} = (x, y, z, w)$. The minus signs are introduced such that all variables $x$, $y$, $z$, and $w$ remain positive in the Euclidean region defined as
\begin{equation}
\label{eq:Euclidean-region}
\mathcal{R}_{\text{E}}
=
\Big\{\, (x, y, z, w) \in \mathbb{R}^4 \;\Big|\; x > 0,\; y > 0,\; z > 0,\; w > 0 \,\Big\}\,.
\end{equation}
Our analytical procedure begins with the following linear basis, collected into $\vec{f}$:
\begin{align}
f_{1} &= \epsilon\, (1 - \epsilon)\, I_{1}\,,
&
f_{2} &= \epsilon^{2} I_{2}\,,
&
f_{3} &= \epsilon^{2} I_{3}\,,
&
f_{4} &= \epsilon^{2} I_{4}\,,
\nonumber \\
f_{5} &= \epsilon^{2} I_{5}\,,
&
f_{6} &= \epsilon^{2} I_{6}\,,
&
f_{7} &= \epsilon^{2} I_{7}\,,
&
f_{8} &= \epsilon^{2} I_{8}\,,
\nonumber \\
f_{9} &= \epsilon^{2} I_{9}\,,
&
f_{10} &= \epsilon^{2} I_{10}\,,
&
f_{11} &= \epsilon^{2} I_{11}\,,
&
f_{12} &= \epsilon^{2} I_{12}\,,
\nonumber \\
f_{13} &= \epsilon^{2} I_{13}\,,
&
f_{14} &= \epsilon^{2} I_{14}\,,
&
f_{15} &= \epsilon^{3}\, (1 - 2\epsilon)\, I_{15}\,,
&
f_{16} &= \epsilon^{2}\, (1 - 2\epsilon)\, I_{16}\,,
\nonumber \\
f_{17} &= \epsilon^{2} I_{17}\,,
&
f_{18} &= \epsilon^{2} I_{18}\,,
&
f_{19} &= \epsilon^{2} I_{19}\,,
&
f_{20} &= \epsilon^{3} I_{20}\,,
\nonumber \\
f_{21} &= \epsilon^{3} I_{21}\,,
&
f_{22} &= \epsilon^{3} I_{22}\,,
&
f_{23} &= \epsilon^{3} I_{23}\,,
&
f_{24} &= \epsilon^{3} I_{24}\,,
\nonumber \\
f_{25} &= \epsilon^{2} I_{25}\,,
&
f_{26} &= \epsilon^{2} I_{26}\,,
&
f_{27} &= \epsilon^{2} I_{27}\,,
&
f_{28} &= \epsilon^{2} I_{28}\,,
\nonumber \\
f_{29} &= \epsilon^{2} I_{29}\,,
&
f_{30} &= \epsilon^{3} I_{30}\,,
&
f_{31} &= \epsilon^{3} I_{31}\,,
&
f_{32} &= \epsilon^{2}\, (1 - 2\epsilon)\, I_{32}\,,
\nonumber \\
f_{33} &= \epsilon^{2}\, (1 - 2\epsilon)\, I_{33}\,,
&
f_{34} &= \epsilon^{4} I_{34}\,,
&
f_{35} &= \epsilon^{4} I_{35}\,,
&
f_{36} &= \epsilon^{4} I_{36}\,,
\nonumber \\
f_{37} &= \epsilon^{4} I_{37}\,,
&
f_{38} &= \epsilon^{3} I_{38}\,,
&
f_{39} &= \epsilon^{3} I_{39}\,,
&
f_{40} &= \epsilon^{3} I_{40}\,,
\\
f_{41} &= \epsilon^{3} I_{41}\,,
&
f_{42} &= \epsilon^{3} I_{42}\,,
&
f_{43} &= \epsilon^{3} I_{43}\,,
&
f_{44} &= \epsilon^{3} I_{44}\,,
\nonumber \\
f_{45} &= \epsilon^{3} I_{45}\,,
&
f_{46} &= \epsilon^{3} I_{46}\,,
&
f_{47} &= \epsilon^{4} I_{47}\,,
&
f_{48} &= \epsilon^{3} I_{48}\,,
\nonumber \\
f_{49} &= \epsilon^{3} I_{49}\,,
&
f_{50} &= \epsilon^{4} I_{50}\,,
&
f_{51} &= \epsilon^{3} I_{51}\,,
&
f_{52} &= \epsilon^{3}\, (1 - 2\epsilon)\, I_{52}\,,
\nonumber \\
f_{53} &= \epsilon^{3}\, (1 - 2\epsilon)\, I_{53}\,,
&
f_{54} &= \epsilon^{3} I_{54}\,,
&
f_{55} &= \epsilon^{3} I_{55}\,,
&
f_{56} &= \epsilon^{4} I_{56}\,,
\nonumber \\
f_{57} &= \epsilon^{4} I_{57}\,,
&
f_{58} &= \epsilon^{3} I_{58}\,,
&
f_{59} &= \epsilon^{3} I_{59}\,,
&
f_{60} &= \epsilon^{3} I_{60}\,,
\nonumber \\
f_{61} &= \epsilon^{3} I_{61}\,,
&
f_{62} &= \epsilon^{4} I_{62}\,,
\nonumber
\end{align}
where the MIs $I_i~ (i = 1, \ldots, 62)$ are displayed in Fig.\ref{fig2}. The differential equations for $\vec{f}$ are derived by differentiation followed by IBP reduction, as detailed in Section \ref{sec:3}. After performing a Magnus transformation on the linear basis $\vec{f}$, the $\mathcal{O}(\epsilon^0)$ terms  in the coefficient matrices of the differential equations are eliminated, thereby yielding the canonical basis $\vec{g}$:
\begin{align}
 \label{eq:gB1}
  g_{1} &= f_{1}\,,
  &
  g_{2} &= x\, f_{2}\,,
  &
  g_{3} &= y\, f_{3}\,,
  \nonumber \\
  g_{4} &= z\, f_{4}\,,
  &
  g_{5} &= w\, f_{5}\,,
  &
  g_{6} &= w r_{1}\, f_{6}\,,
  \nonumber \\
  g_{7} &= x\, f_{7}\,,
  &
  g_{8} &= y\, f_{8}\,,
  &
  g_{9} &= z\, f_{9}\,,
  \nonumber \\
  g_{10} &= w\, f_{10}\,,
  &
  g_{11} &= 2\, f_{7} + (1+x)\, f_{11}\,,
  &
  g_{12} &= 2\, f_{8} + (1+y)\, f_{12}\,,
  \nonumber \\
  g_{13} &= 2\, f_{9} + (1+z)\, f_{13}\,,
  &
  g_{14} &= 2\, f_{10} + (1+w)\, f_{14}\,,
  &
  g_{15} &= w\, f_{15}\,,
  \nonumber \\
  g_{16} &= {\textstyle r_{1}/(4+w) \sum_{i} \alpha_{16,i}\, f_{i}}\,,
  &
  g_{17} &= y r_{1}\, f_{17}\,,
  &
  g_{18} &= z r_{1}\, f_{18}\,,
  \nonumber \\
  g_{19} &= x r_{1}\, f_{19}\,,
  &
  g_{20} &= r_{2}\, f_{20}\,,
  &
  g_{21} &= r_{1} r_{2}\, f_{21}\,,
  \nonumber \\
  g_{22} &= (y-w)\, f_{22}\,,
  &
  g_{23} &= r_{2}\, f_{23}\,,
  &
  g_{24} &= r_{2}\, f_{24}\,,
  \nonumber \\
  g_{25} &= (y-w)\, f_{25}\,,
  &
  g_{26} &= r_{2}\, f_{26}\,,
  &
  g_{27} &= r_{2}\, f_{27}\,,
  \nonumber \\
  g_{28} &\rlap{$\displaystyle {} = 3\, (x-z+w)\, f_{23} + 2\, (x-z+w)\, f_{26} + 2 x w\, f_{28}\,,$}
  &
  &
  \\
  g_{29} &\rlap{$\displaystyle {} = 3\, (x+z-w)\, f_{24} + 2\, (x+z-w)\, f_{27} + 2 x z\, f_{29}\,,$}
  &
  &
  &
  &
  \nonumber \\
  g_{30} &= (y-w)\, f_{30}\,,
  &
  g_{31} &= r_{2}\, f_{31}\,,
  &
  g_{32} &= {\textstyle r_{1}/w \sum_{i} \alpha_{32,i}\, f_{i}}\,,
  \nonumber \\
  g_{33} &= {\textstyle r_{1}/w \sum_{i} \alpha_{33,i}\, f_{i}}\,,
  &
  g_{34} &= (y-w)\, f_{34}\,,
  &
  g_{35} &= (y-w)\, f_{35}\,,
  \nonumber \\
  g_{36} &= r_{2}\, f_{36}\,,
  &
  g_{37} &= r_{2}\, f_{37}\,,
  &
  g_{38} &= y\, (y-w)\, f_{38}\,,
  \nonumber \\
  g_{39} &= z r_{2}\, f_{39}\,,
  &
  g_{40} &= x r_{2}\, f_{40}\,,
  &
  g_{41} &= (y-w)\, r_{1}\, f_{41}\,,
  \nonumber \\
  g_{42} &= r_{1} r_{2}\, f_{42}\,,
  &
  g_{43} &= r_{1} r_{2}\, f_{43}\,,
  &
  g_{44} &= 2 w\, f_{35} - y\, f_{44}\,,
  \nonumber \\
  g_{45} &= {\textstyle \sum_{i} \alpha_{45,i}\, f_{i}}\,,
  &
  g_{46} &= {\textstyle \sum_{i} \alpha_{46,i}\, f_{i}}\,,
  &
  g_{47} &= w r_{2}\, f_{47}\,,
  \nonumber \\
  g_{48} &= x y\, f_{48}\,,
  &
  g_{49} &= x y r_{1}\, f_{49}\,,
  &
  g_{50} &= (x+y-z)\, f_{50}\,,
  \nonumber \\
  g_{51} &= (x+y-z+x y)\, f_{51}\,,
  &
  g_{52} &= x\, ( f_{52} - f_{54} )\,,
  &
  g_{53} &= (y-z)\, ( f_{53} - f_{55} )\,,
  \nonumber \\
  g_{54} &= x y\, f_{54}\,,
  &
  g_{55} &= x y\, f_{55}\,,
  &
  g_{56} &= x\, (y-w)\, f_{56}\,,
  \nonumber \\
  g_{57} &= [\, x y - (y-z) w \,]\, f_{57}\,,
  &
  g_{58} &\rlap{$\displaystyle {} = x\, [\, 2 y\, f_{56} + (w+2y+wy)\, f_{58} + (2 y-w)\, f_{60} \,]\,,$}
  &
  &
  \nonumber \\
  g_{59} &\rlap{$\displaystyle {} = 2w\, (y-z)\, f_{57} + [\, w\, (y-z) + x y\, (2+w) \,]\, f_{59} + [\, 2 x y - w\, (y-z) \,]\, f_{61}\,,$}
  &
  &
  &
  &
  \nonumber \\
  g_{60} &\rlap{$\displaystyle {} = x r_{1}\, [\, 2\, f_{56} + (1+y)\, f_{58} + f_{60} \,]\,,$}
  &
  &
  \nonumber \\
  g_{61} &\rlap{$\displaystyle {} = r_{1}\, [\, 2\, (y-z)\, f_{57} + (y-z+x y)\, f_{59} + (y-z)\, f_{61} \,]\,, $}
  &
  &
  &
  &
  \nonumber \\
  g_{62} &\rlap{$\displaystyle {} = w\, [\, x\, f_{56} + (y-z)\, f_{57} + x y\, f_{62} \,]\,,$}
  &
  &
  \nonumber
\end{align}
where the non-vanishing coefficients $\alpha_{i,j}$ are given in Appendix \ref{sec:appA}, and the square roots $r_1$ and $r_2$ are defined by
\begin{equation}
r_1^2 = w\, (4 + w)\,,
\qquad\quad
r_2^2 = \lambda(x,\, z,\, w)\,,
\end{equation}
with $\lambda$ the K\"all\'en function,
\begin{equation}
\lambda(x,\, y,\, z) = x^2 + y^2 + z^2 - 2 x y - 2 y z - 2 z x\,.
\end{equation}
The differential one-form matrix corresponding to the canonical basis $\vec{g}$ is obtained via the similarity transformation in Eq.\eqref{eq:dA-DEs}; this matrix, in turn, determines the alphabet, comprising $24$ even and $12$ odd letters \cite{matijasic2024}:
\begin{itemize}
\item even letters
\begin{align}
\omega_{1} &= x\,,
&
\omega_{9} &= 4+w\,,
&
\omega_{17} &= (1+z)\, (1+w)-x\,,
\nonumber
\\
\omega_{2} &= y\,,
&
\omega_{10} &= y-z\,,
&
\omega_{18} &= (1+x)\, (1+w)-z\,,
\nonumber
\\
\omega_{3} &= z\,,
&
\omega_{11} &= y-w\,,
&
\omega_{19} &= xy-(y-z)\,w\,,
\\
\omega_{4} &= w\,,
&
\omega_{12} &= 1-y+w\,,
&
\omega_{20} &= xy+(y-z)\, (y-w)\,,
\nonumber
\\
\omega_{5} &= 1+x\,,
&
\omega_{13} &= y-z+x\,,
&
\omega_{21} &= (x-z)^{2}-w\, (1+x)\, (1+z)\,,
\nonumber
\\
\omega_{6} &= 1+y\,,
&
\omega_{14} &= y-z+xy\,,
&
\omega_{22} &= r_2^2\,,
\nonumber
\\
\omega_{7} &= 1+z\,,
\quad
&
\omega_{15} &= y-z+x\,(1+y)\,,
\quad
&
\omega_{23} &= (x+y-z)^{2} + xw\, (y-z)\,,
\nonumber
\\
\omega_{8} &= 1+w\,,
&
\omega_{16} &= y^{2}-w\,(1+y)\,,
&
\omega_{24} &= x^{2} y^{2}-w\,(y-z)\, (xy+y-z)\,,
\nonumber
\end{align}
\item odd letters
\begin{align}
\omega_{25} &= \frac{w+r _{1}}{w-r _{1}}\,,
&
\omega_{31} &= \frac{x-z+w+r _{2}}{x-z+w-r _{2}}\,,
\nonumber
\\
\omega_{26} &= \frac{2y-w+r _{1}}{2y-w-r _{1}}\,,
&
\omega_{32} &= \frac{2-x+z+w+r _{2}}{2-x+z+w-r _{2}}\,,
\nonumber
\\
\omega_{27} &= \frac{w\,(2+x+z)+(x-z)\, r _{1}}{w\,(2+x+z)-(x-z)\, r _{1}}\,,
&
\omega_{33} &= \frac{2+x-z+w+r _{2}}{2+x-z+w-r _{2}}\,,
\\
\omega_{28} &= \frac{2\, (y-z)+x\, (2+w+ r _{1})}{2\, (y-z)+x\, (2+w- r _{1})}\,,
&
\omega_{34} &= \frac{x+2y-z-w+r _{2}}{x+2y-z-w-r _{2}}\,,
\nonumber
\\
\omega_{29} &= \frac{2w\,(y-z)+xy\, (w+r _{1})}{2w\,(y-z)+xy\, (w-r _{1})}\,,
\qquad
&
\omega_{35} &= \frac{(1-z)\, (x-z)-(1+z)\, (w-r _{2})}{(1-z)\, (x-z)-(1+z)\, (w+r _{2})}\,,
\nonumber
\\
\omega_{30} &= \frac{x-z-w+r _{2}}{x-z-w-r _{2}}\,,
&
\omega_{36} &= \frac{w\, (2-x-z+w)+r _{1} r _{2}}{w\, (2-x-z+w)-r _{1} r _{2}}\,.
\nonumber
\end{align}
\end{itemize}
Expressing the iterative solutions of the canonical basis in terms of GPLs typically requires rationalizing the square roots in the alphabet. For $\mathcal{B}_1$, although both $r_1$ and $r_2$ can be simultaneously rationalized by an appropriate transformation, this is unnecessary, as not all MIs depend on both square roots. A more efficient strategy is therefore to treat subsystems characterized by distinct radical structures separately and rationalize each subsystem individually. Next, we demonstrate how to determine the dependence of each MI in a given family branch on a specified set of square roots, which serves as the basis for constructing the corresponding subsystems.

\par
We introduce the functional $\mathcal{Z}_{r_1,\ldots,r_N}$, referred to as the radical index, an $N$-bit binary number that encodes the dependence of a function on the set of independent square roots $\{r_1, \ldots, r_N\}$. It is defined as
\begin{equation}
\mathcal{Z}_{r_1,\ldots,r_N}\left[\, h \,\right] = (\mu_1 \ldots \mu_N)
\quad~ \text{with} \quad~
\mu_i
=
\begin{cases}
\;0\,, & \text{if} \quad h\,|_{\,r_i \; \longrightarrow\; -r_i} = h
\\
\;1\,, & \text{if} \quad h\,|_{\,r_i \; \longrightarrow\; -r_i} \neq h
\end{cases}
\end{equation}
For a composite expression, the dependence on each root accumulates as the constituent components are combined. This accumulated dependence is captured by the binary operation \textit{bitwise OR}, denoted by $|$, which is defined as follows:
\begin{equation}
(\mu_1 \ldots \mu_N) \,\big|\, (\nu_1 \ldots \nu_N) = (\lambda_1 \ldots \lambda_N)\,,
\quad~ \text{where} \quad~
\lambda_i = \max\{\mu_i\,, \nu_i\}\,.
\end{equation}
It then follows that
\begin{equation}
\mathcal{Z}_{r_1,\ldots,r_N}\left[\, h_1 \smile h_2 \,\right]
\leqslant
\mathcal{Z}_{r_1,\ldots,r_N}\left[\, h_1 \,\right] \,\big|\, \mathcal{Z}_{r_1,\ldots,r_N}\left[\, h_2 \,\right]\,,
\end{equation}
where $h_1 \smile h_2$ represents any fundamental arithmetic operation on $h_1$ and $h_2$, i.e., addition, subtraction, multiplication, or division. It is noted that the equality generally holds, except in cases where cancellations occur between square roots, including degenerate instances where either $h_1$ or $h_2$ vanishes.

\par
\begin{figure}[htbp]
\centering
\begin{minipage}{0.95\textwidth}
\includegraphics[width=1.0\textwidth]{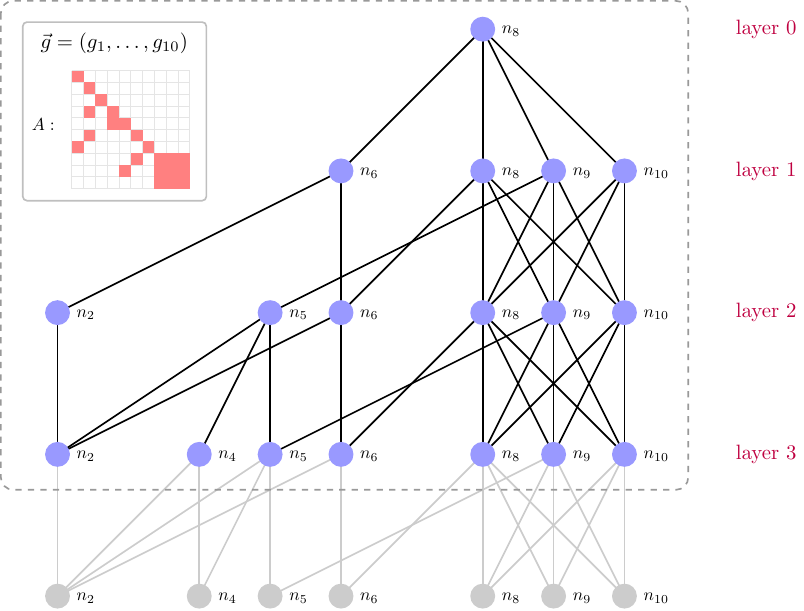}
\end{minipage}
\caption{
An example of a tropical tree rooted at $g_8$. This canonical differential system involves ten MIs $g_i~ (i = 1, 2, \ldots, 10)$, with the nonzero elements of $d\mathbbm{A}$ highlighted in red in the top-left panel. No new MIs are generated beyond layer 3, and accordingly, the minimal subsystem containing $g_8$ is $\texttt{Node}\textsf{(}g_8\textsf{)} = \big\{g_2,\, g_4,\, g_5,\, g_6,\, g_8,\, g_9,\, g_{10}\big\}$.
}
\label{fig3}
\end{figure}

The zeroth-order MIs are constants, and as such, their radical indices are zero. According to the recursion relation \eqref{eq:recursion-relation}, the radical index of a canonical MI at any order is determined by performing a bitwise OR on the radical indices of all nonzero elements of $d\mathbbm{A}$ involved in the iteration, as well as those of the corresponding nonzero MIs from the preceding order, namely
\begin{equation}
\label{eq:recursiveZ}
\mathcal{Z}_{r_1,\ldots,r_N}\big[\, g_i^{(p)} \,\big]
=
\Big|_{\,j:\; d\mathbbm{A}_{ij} g_j^{(p-1)}}
\Big(\,
\mathcal{Z}_{r_1,\ldots,r_N}\big[\, d\mathbbm{A}_{ij} \,\big]
\,\big|\,
\mathcal{Z}_{r_1,\ldots,r_N}\big[\, g_j^{(p-1)} \,\big]
\,\Big)\,.
\end{equation}
Equation \eqref{eq:recursiveZ} defines a tree-like structure for efficiently computing the radical index for each canonical MI. Since loops are permitted in this structure, we refer to it as a tropical tree. Each node in a tropical tree represents a canonical MI, with each edge corresponding to a nonzero element of $d\mathbbm{A}$. The tree expands layer by layer, with each layer of nodes simultaneously generating new nodes via the edges, continuing until no additional distinct nodes are generated. We denote the tropical tree rooted at $g_i$ as \texttt{Tree}\textsf{($g_i$)}, and the set of all its nodes as \texttt{Node}\textsf{($g_i$)}. Evidently, \texttt{Node}\textsf{($g_i$)} defines the minimal subsystem that contains the MI $g_i$, with the radical index of $g_i$ given by
\begin{equation}
\mathcal{Z}_{r_1,\ldots,r_N}\left[\, g_i \,\right]
=
\Big|_{\,j:\; g_j \in\, \texttt{Node}\text{($g_i$)}}
\Big(\,
\Big|_{\,k:\; d\mathbbm{A}_{jk} \neq 0}
\,
\mathcal{Z}_{r_1,\ldots,r_N}\big[\, d\mathbbm{A}_{jk} \,\big]
\,\Big)\,.
\end{equation}
The radical index of any MI in this subsystem is at most that of $g_i$. Figure \ref{fig3} illustrates a tropical tree rooted at a MI of a toy canonical differential system. In this example, the tree stabilizes after three layers from the root integral, with no new MIs introduced beyond this depth.

\par
The canonical MIs of the $\mathcal{B}_1$ branch are classified into four disjoint subsets according to their radical structures, defined as
\begin{equation}
S_{1\,(\mu\nu)}
=
\big\{\, g_i \,\big|\, i = 1, 2, \ldots, 62\,,\, \mathcal{Z}_{r_1,r_2}\left[\, g_i \,\right] = (\mu\nu) \,\big\}\,,
\qquad
\mu\nu = 00,\, 10,\, 01, ~\text{and} ~11\,,
\end{equation}
with the elements of each subset collected in Table \ref{tab1}. Based on these four subsets, we define four distinct subsystems within the $\mathcal{B}_1$ branch, as follows:
\begin{equation}
\mathcal{B}_{1\,(\mu\nu)}
=
\bigcup_{g\, \in S_{1\,(\mu\nu)}}
\texttt{Node}\textsf{($g$)}\,,
\qquad\quad
(\mu\nu = 00,\, 10,\, 01,\, 11)\,.
\end{equation}
To make the analytic expressions of the MIs as compact as possible, those with radical index $(\mu\nu)$ are derived from the solution of the subsystem $\mathcal{B}_{1\,(\mu\nu)}$. The subsystem $\mathcal{B}_{1\,(00)}$ is the simplest, as its system of differential equations features a coefficient matrix that is rational in $\vec{x}$. The subsystems $\mathcal{B}_{1\,(10)}$ and $\mathcal{B}_{1\,(01)}$ each involve only the square roots $r_1$ and $r_2$, and can be rationalized via the variable transformations
\begin{equation}
\label{eq:phi1}
\Phi_1(w) \colon~  w ~\longmapsto~ \mathbbm{w}\,,
\qquad
w = \frac{\mathbbm{w}^2}{1 + \mathbbm{w}}\,,
\end{equation}
and
\begin{equation}
\label{eq:phi2}
\Phi_2(x, z; w) \colon ~ (x, z) ~\longmapsto~ (\mathbbm{x}, \mathbbm{z})\,,
\qquad
\left\lbrace\,
\begin{aligned}
&  x = \mathbbm{x}\, (1 - \mathbbm{z})\, w\,, \\
&  z = \mathbbm{z}\, (1 - \mathbbm{x})\, w\,,
\end{aligned}
\right.
\end{equation}
respectively. For the most intricate subsystem $\mathcal{B}_{1\,(11)}$, involving both square roots, the variable transformation $\Psi$ that simultaneously rationalizes $r_1$ and $r_2$ is the composition of $\Phi_1$ and $\Phi_2$,
\begin{equation}
\label{eq:Psi}
\Psi(x, z, w) = \Phi_1(w) \,\circ\, \Phi_2(x, z; w)\,.
\end{equation}
Upon applying the rationalization transformation, the two square roots take the form of
\begin{equation}
\label{eq:Rr12}
r_1 = \frac{\mathbbm{w}\, (2 + \mathbbm{w})}{1 + \mathbbm{w}}\,,
\qquad\quad
r_2 = (1 - \mathbbm{x} - \mathbbm{z})\, \frac{\mathbbm{w}^2}{1 + \mathbbm{w}}\,,
\end{equation}
and the corresponding MIs can be expressed in terms of GPLs.
\begin{table}[htbp]
\footnotesize
\setlength{\tabcolsep}{6pt}
\centering
\renewcommand{\arraystretch}{1.08}
\begin{tabular}{
!{\vrule width 1.2pt}
p{1.4cm}<{\centering}
|
p{3.2cm}<{\centering}
p{7.8cm}<{\centering}
p{1.1cm}<{\centering}
!{\vrule width 1.2pt}
}
\Hline
\textsf{Subset} & \textsf{Radical structure} & \textsf{Master integrals~ ($\boldsymbol{g_i}$)} & \textsf{Count} \\ \Hline
$S_{1\,(00)}$ & $\{\;\}$ & $1\mbox{-}5, 7\mbox{-}14, 22, 25$ & 15 \\
$S_{1\,(10)}$ & $\{r_1\}$ & $6, 15\mbox{-}19, 30, 32, 34, 35, 38, 41, 44$ & 13 \\
$S_{1\,(01)}$ & $\{r_2\}$ & $20, 23, 24, 26\mbox{-}29, 48, 52\mbox{-}55$ & 12 \\
$S_{1\,(11)}$ & $\{r_1, r_2\}$ & $21, 31, 33, 36, 37, 39, 40, 42, 43, 45\mbox{-}47, 49\mbox{-}51, 56\mbox{-}62$ & 22 \\ \Hline
\end{tabular}
\caption{Classification of the MIs in $\mathcal{B}_1$ according to their radical structures.}
\label{tab1}
\end{table}

\par
To determine the solution to the system of differential equations, it is essential to specify the boundary conditions. These can be obtained by expansion by regions \cite{Smirnov:1994tg,Beneke:1997zp}, by matching analytic constants at specific kinematic points using the numerical \texttt{PSLQ} algorithm \cite{ferguson1999}, or by direct evaluation in appropriate limits. For a canonical basis, the system of differential equations is typically Fuchsian, with singularities located at the zeros of the letters. These singularities encompass both the genuine Landau singularities of the MIs and, potentially, spurious singularities. We employ the method of expansion by regions, implemented in \texttt{asy2.m} \cite{Pak:2010pt,Jantzen:2011nz,Jantzen:2012mw}, to analyze the asymptotic behavior of the integrals near singularities. At spurious singularities, the regularity condition requires that the coefficients of the corresponding logarithmically divergent terms in the iterated solution vanish. Furthermore, when the integral converges to zero at a given point, a stronger boundary condition is imposed, referred to as the vanishing condition. For the $\mathcal{B}_1$ branch, consisting of $62$ MIs, there are at most $61$ linearly homogeneous and independent boundary conditions, which are selected as follows:
\begin{itemize}
\item regularity conditions (60)
        \begin{itemize}
        \item regularity at $x = 0$: $g_{7,\,24,\,50,\,51,\,53,\,57,\,59,\,61}$
        \item regularity at $w = 0$: $g_{6,\,10,\,15,\,21,\,23,\,30,\,31,\,35,\,36,\,37,\,41,\,42,\,43,\,56,\,57,\,60,\,61,\,62}$
        \item regularity at $y - z = 0$: $g_{48,\,49,\,50,\,51,\,53,\,57}$
        \item regularity at $y - w = 0$: $g_{22,\,25,\,30,\,35,\,38,\,41,\,48,\,49,\,52,\,58,\,62}$
        \item regularity at $1 - y + w = 0$: $g_{25}$
        \item regularity at $(1 + z)\, (1 + w) - x = 0$: $g_{26}$
        \item regularity at $x y + (y - z)\,(y - w) = 0$: $g_{48,\,49,\,52,\,53,\,54,\,56,\,60,\,62}$
        \item regularity at $\lambda(x, z, w) = 0$: $g_{20,\,21,\,23,\,26,\,27,\,36,\,37}$
        \end{itemize}
\item vanishing condition (1)
        \begin{itemize}
        \item vanishing at $y - w = 0$: $g_{34}$
        \end{itemize}
\end{itemize}
Given that the boundary conditions outlined above are homogeneous, an additional non-homogeneous boundary condition or non-zero boundary constant must be introduced, which can be obtained from the analytic expression of $g_1$,
\begin{equation}
g_1 = 1 + \frac{\pi^2}{2}\,\epsilon^2 - \frac{8}{3}\zeta(3)\, \epsilon^3 + \frac{7\pi^4}{40}\, \epsilon^4 + \mathcal{O}(\epsilon^5)\,.
\end{equation}

\par
In Appendix \ref{sec:appB}, we present the explicit expressions for the 62 canonical MIs of $\mathcal{B}_1$ up to $\mathcal{O}(\epsilon^2)$. The analytic expressions up to $\mathcal{O}(\epsilon^4)$ are provided in the supplementary file \texttt{B1.m}.

\section{Family branch $\mathcal{B}_{2}$}
\label{sec:5}
\par
\begin{figure}[htbp]
\centering
\begin{minipage}{0.95\textwidth}
\includegraphics[width=1.0\textwidth]{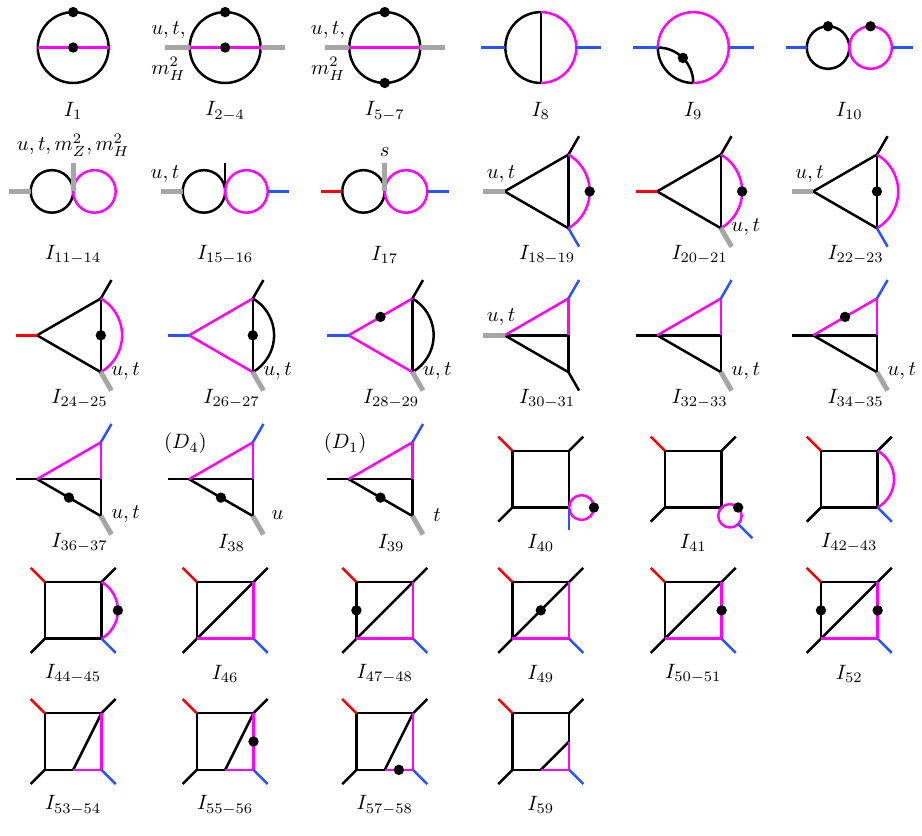}
\end{minipage}
\caption{Same as Fig.\ref{fig2}, but for $\mathcal{B}_{2}$.}
\label{fig4}
\end{figure}

For the $\mathcal{B}_2$ branch, the top-topology exhibits symmetry under the exchange $u \leftrightarrow t$. Therefore, it is convenient to choose $\big\{u,\, t,\, m_Z^2,\, m_H^2,\, m_W^2\big\}$ as the independent kinematic invariants for the Feynman integrals in $\mathcal{B}_2$. Taking $m_W$ as the characteristic scale, the dimensionless Feynman integrals are expressed in terms of the following variables:
\begin{equation}
x = -\, \frac{u}{m_W^2}\,,
\qquad
y = -\, \frac{t}{m_W^2}\,,
\qquad
z = -\, \frac{m_Z^2}{m_W^2}\,,
\qquad
w = -\, \frac{m_H^2}{m_W^2}\,.
\end{equation}
We choose the following linear basis for the $\mathcal{B}_2$ branch,
\begin{align}
f_{1} &= \epsilon\, (1 - \epsilon)\, I_{1}\,,
&
f_{2} &= \epsilon^{2} I_{2}\,,
&
f_{3} &= \epsilon^{2} I_{3}\,,
&
f_{4} &= \epsilon^{2} I_{4}\,,
\nonumber \\
f_{5} &= \epsilon^{2} I_{5}\,,
&
f_{6} &= \epsilon^{2} I_{6}\,,
&
f_{7} &= \epsilon^{2} I_{7}\,,
&
f_{8} &= \epsilon^{3}\, (1 - 2\epsilon)\, I_{8}\,,
\nonumber \\
f_{9} &= \epsilon^{2}\, (1 - 2\epsilon)\, I_{9}\,,
&
f_{10} &= \epsilon^{2} I_{10}\,,
&
f_{11} &= \epsilon^{2} I_{11}\,,
&
f_{12} &= \epsilon^{2} I_{12}\,,
\nonumber \\
f_{13} &= \epsilon^{2} I_{13}\,,
&
f_{14} &= \epsilon^{2} I_{14}\,,
&
f_{15} &= \epsilon^{2} I_{15}\,,
&
f_{16} &= \epsilon^{2} I_{16}\,,
\nonumber \\
f_{17} &= \epsilon^{2} I_{17}\,,
&
f_{18} &= \epsilon^{3} I_{18}\,,
&
f_{19} &= \epsilon^{3} I_{19}\,,
&
f_{20} &= \epsilon^{3} I_{20}\,,
\nonumber \\
f_{21} &= \epsilon^{3} I_{21}\,,
&
f_{22} &= \epsilon^{3} I_{22}\,,
&
f_{23} &= \epsilon^{3} I_{23}\,,
&
f_{24} &= \epsilon^{3} I_{24}\,,
\nonumber \\
f_{25} &= \epsilon^{3} I_{25}\,,
&
f_{26} &= \epsilon^{3} I_{26}\,,
&
f_{27} &= \epsilon^{3} I_{27}\,,
&
f_{28} &= \epsilon^{2}\, (1 - 2\epsilon)\, I_{28}\,,
\nonumber \\
f_{29} &= \epsilon^{2}\, (1 - 2\epsilon)\, I_{29}\,,
&
f_{30} &= \epsilon^{4} I_{30}\,,
&
f_{31} &= \epsilon^{4} I_{31}\,,
&
f_{32} &= \epsilon^{4} I_{32}\,,
\nonumber \\
f_{33} &= \epsilon^{4} I_{33}\,,
&
f_{34} &= \epsilon^{3} I_{34}\,,
&
f_{35} &= \epsilon^{3} I_{35}\,,
&
f_{36} &= \epsilon^{3} I_{36}\,,
\\
f_{37} &= \epsilon^{3} I_{37}\,,
&
f_{38} &= \epsilon^{3} I_{38}\,,
&
f_{39} &= \epsilon^{3} I_{39}\,,
&
f_{40} &= \epsilon^{3} I_{40}\,,
\nonumber \\
f_{41} &= \epsilon^{3} I_{41}\,,
&
f_{42} &= \epsilon^{3}\, (1 - 2\epsilon)\, I_{42}\,,
&
f_{43} &= \epsilon^{3}\, (1 - 2\epsilon)\, I_{43}\,,
&
f_{44} &= \epsilon^{3} I_{44}\,,
\nonumber \\
f_{45} &= \epsilon^{3} I_{45}\,,
&
f_{46} &= \epsilon^{4} I_{46}\,,
&
f_{47} &= \epsilon^{3} I_{47}\,,
&
f_{48} &= \epsilon^{3} I_{48}\,,
\nonumber \\
f_{49} &= \epsilon^{3} I_{49}\,,
&
f_{50} &= \epsilon^{3} I_{50}\,,
&
f_{51} &= \epsilon^{3} I_{51}\,,
&
f_{52} &= \epsilon^{2} I_{52}\,,
\nonumber \\
f_{53} &= \epsilon^{4} I_{53}\,,
&
f_{54} &= \epsilon^{4} I_{54}\,,
&
f_{55} &= \epsilon^{3} I_{55}\,,
&
f_{56} &= \epsilon^{3} I_{56}\,,
\nonumber \\
f_{57} &= \epsilon^{3} I_{57}\,,
&
f_{58} &= \epsilon^{3} I_{58}\,,
&
f_{59} &= \epsilon^{4} I_{59}\,,
\nonumber
\end{align}
where $\{ I_i \,|\, i=1, \ldots, 59 \}$, a set of MIs for $\mathcal{B}_2$, is illustrated in Fig.\ref{fig4}. The canonical basis, constructed using the Magnus transformation, is explicitly given by
\begin{align}
\label{eq:gB2}
g_{1} &= f_{1}\,,
&
g_{2} &= x\, f_{2}\,,
&
g_{3} &= y\, f_{3}\,,
\nonumber \\
g_{4} &= w\, f_{4}\,,
&
g_{5} &= 2\, f_{2}+(1+x)\, f_{5}\,,
&
g_{6} &= 2\, f_{3}+(1+y)\, f_{6}\,,
\nonumber \\
g_{7} &= 2\, f_{4}+(1+w)\, f_{7}\,,
&
g_{8} &= w\, f_{8}\,,
&
g_{9} &= r_{1} / (4 + w)\, {\textstyle \sum_{i} \beta_{9,i}\, f_{i}}\,,
\nonumber \\
g_{10} &= w r_{1}\, f_{10}\,,
&
g_{11} &= x\, f_{11}\,,
&
g_{12} &= y\, f_{12}\,,
\nonumber \\
g_{13} &= z\, f_{13}\,,
&
g_{14} &= w\, f_{14}\,,
&
g_{15} &= x r_{1}\, f_{15}\,,
\nonumber \\
g_{16} &= y r_{1}\, f_{16}\,,
&
g_{17} &= z r_{1}\, f_{17}\,,
&
g_{18} &= (x-w)\, f_{18}\,,
\nonumber \\
g_{19} &= (y-w)\, f_{19}\,,
&
g_{20} &= (x-z)\, f_{20}\,,
&
g_{21} &= (y-z)\, f_{21}\,,
\nonumber \\
g_{22} &= (x-w)\, f_{22}\,,
&
g_{23} &= (y-w)\, f_{23}\,,
&
g_{24} &= (x-z)\, f_{24}\,,
\nonumber \\
g_{25} &= (y-z)\, f_{25}\,,
&
g_{26} &= (x-w)\, f_{26}\,,
&
g_{27} &= (y-w)\, f_{27}\,,
\nonumber \\
g_{28} &= r_{1} / w\, {\textstyle \sum_{i} \beta_{28,i}\, f_{i}}\,,
&
g_{29} &= r_{1} / w\, {\textstyle \sum_{i} \beta_{29,i}\, f_{i}}\,,
&
g_{30} &= (x-w)\, f_{30}\,,
\nonumber \\
g_{31} &= (y-w)\, f_{31}\,,
&
g_{32} &= (x-w)\, f_{32}\,,
&
g_{33} &= (y-w)\, f_{33}\,,
\\
g_{34} &= (x-w)\, r_{1}\, f_{34}\,,
&
g_{35} &= (y-w)\, r_{1}\, f_{35}\,,
&
g_{36} &= x\, (x-w)\, f_{36}\,,
\nonumber \\
g_{37} &= y\, (y-w)\, f_{37}\,,
&
g_{38} &= 2 w\, f_{32}-x\, f_{38}\,,
&
g_{39} &= 2 w\, f_{33}-y\, f_{39}\,,
\nonumber \\
g_{40} &= (x y-z w)\, f_{40}\,,
&
g_{41} &= (x y-z w)\, r_{1}\, f_{41}\,,
&
g_{42} &= (x-z)\, ( f_{42} - f_{44} )\,,
\nonumber \\
g_{43} &= (y-z)\, ( f_{43} - f_{45} )\,,
&
g_{44} &= (x y-z w)\, f_{44}\,,
&
g_{45} &= (x y-z w)\, f_{45}\,,
\nonumber \\
g_{46} &= r_{2}\, f_{46}\,,
&
g_{47} &= z\, (y-w)\, f_{47}\,,
&
g_{48} &= z\, (x-w)\, f_{48}\,,
\nonumber \\
g_{49} &= r_{3}\, f_{49}\,,
&
g_{50} &= r_{4}\, [\, f_{50} + f_{51} / (1+\mathbbm{w}) \,]\,,
&
g_{51} &= r_{5}\, [\, f_{50} / (1+\mathbbm{w}) + f_{51} \,]\,,
\nonumber \\
g_{52} &= 1 / w\, {\textstyle \sum_{i} \beta_{52,i}\, f_{i}}\,,
&
g_{53} &= x\, (y-w)\, f_{53}\,,
&
g_{54} &= y\, (x-w)\, f_{54}\,,
\nonumber \\
g_{55} &\rlap{$\displaystyle {} = r_{1}\, [\, 2\, (x-z)\, f_{53} + (x-z+x y - z w)\, f_{55} + (x-z)\, f_{57} \,]\,, $}
&
&
&
&
\nonumber \\
g_{56} &\rlap{$\displaystyle {} = r_{1}\, [\, 2\, (y-z)\, f_{54} + (y-z+x y - z w)\, f_{56} + (y-z)\, f_{58} \,]\,, $}
&
&
&
&
\nonumber \\
g_{57} &\rlap{$\displaystyle {} = 2 w\, (x-z)\, f_{53} + [\, w\, (x-z) + (2+w)\, (x y - z w) \,]\, f_{55} + [\, x\, (y-w) + (x y - z w) \,]\, f_{57}\,, $}
&
&
&
&
\nonumber \\
g_{58} &\rlap{$\displaystyle {} = 2 w\, (y-z)\, f_{54} + [\, w\, (y-z) + (2+w)\, (x y - z w) \,]\, f_{56} + [\, y\, (x-w) + (x y - z w) \,]\, f_{58}\,, $}
&
&
&
&
\nonumber \\
g_{59} &\rlap{$\displaystyle {} = w\, [\, (x-z)\, f_{53} + (y-z)\, f_{54} + (x y-z w)\, f_{59} \,]\,,$}
\nonumber
\end{align}
with the non-vanishing coefficients $\beta_{i,j}$ given in Appendix \ref{sec:appA}. The five square roots, $r_{1\mbox{-}5}$, appearing in the canonical basis, are given by
\begin{align}
\label{eq:B2-sqrts}
&
r_1^2 = w\, (4 + w)\,,
&&
r_4^2 = \lambda(\mathbbm{w} - x,\, \mathbbm{w} z,\, \mathbbm{w} + y + \mathbbm{w} y)\,,
\nonumber
\\
&
r_2^2 = \lambda(z,\, w,\, z + w - x - y)\,,
&&
r_5^2 = \lambda(\mathbbm{w} - y,\, \mathbbm{w} z,\, \mathbbm{w} + x + \mathbbm{w} x)\,,
\\
&r_3^2 = \lambda(1 + x - z,\, 1 + y - z,\, z w - x y)\,,
&&
\nonumber
\end{align}
where $\mathbbm{w}$ is defined in Eq.\eqref{eq:phi1}. It is worth noting that $r_4$ and $r_5$ are nested square roots of $\vec{x}$, signifying a more intricate structure for the $\mathcal{B}_2$ branch compared to the $\mathcal{B}_1$ branch. The differential one-form matrix for the canonical basis is constructed as described in Eq.\eqref{eq:dA-DEs}, yielding an alphabet consisting of $26$ even and $27$ odd letters,
\begin{itemize}
\item even letters
\begin{align}
\omega_{1} &= x\,,
&
\omega_{10} &= y-z\,,
&
\omega_{19} &= r_{2}^{2}\,,
\nonumber
\\
\omega_{2} &= y\,,
&
\omega_{11} &= x-w\,,
&
\omega_{20} &= r_{3}^{2}\,,
\nonumber
\\
\omega_{3} &= z\,,
&
\omega_{12} &= y-w\,,
&
\omega_{21} &= x^{2}-w\,(1+x)\,,
\nonumber
\\
\omega_{4} &= w\,,
&
\omega_{13} &= 1+x-z\,,
&
\omega_{22} &= y^{2}-w\,(1+y)\,,
\nonumber
\\
\omega_{5} &= 1+x\,,\quad
&
\omega_{14} &= 1+y-z\,,\quad
&
\omega_{23} &= x\,(1+y) - z\,(1+w)\,,
\\
\omega_{6} &= 1+y\,,
&
\omega_{15} &= 1-x+w\,,
&
\omega_{24} &= y\,(1+x) - z\,(1+w)\,,
\nonumber
\\
\omega_{7} &= 1+w\,,
&
\omega_{16} &= 1-y+w\,,
&
\omega_{25} &= w\, (x-z)^2 - x\, (y-w)\, (xy-zw)\,,
\nonumber
\\
\omega_{8} &= 4+w\,,
&
\omega_{17} &= x+y-z-w\,,
&
\omega_{26} &= w\, (y-z)^2 - y\, (x-w)\, (xy-zw)\,.
\nonumber
\\
\omega_{9} &= x-z\,,
&
\omega_{18} &= x y-z w\,,
\nonumber
\end{align}
\item odd letters
\begin{align}
\omega_{27} &= \frac{w+r_{1}}{w-r_{1}}\label{eq:t2-odd}\,,
&
\omega_{41} &= \frac{x+y+\mathbbm{w}\, (y-z)+r_{4}}{x+y+\mathbbm{w}\, (y-z)-r_{4}}\,,
\nonumber
\\
\omega_{28} &= \frac{2 x-w+r_{1}}{2 x-w-r_{1}}\,,
&
\omega_{42} &= \frac{x-y- \mathbbm{w}\, (y-z)+r_{4}}{x-y- \mathbbm{w}\, (y-z)-r_{4}}\,,
\nonumber
\\
\omega_{29} &= \frac{2 y-w+r_{1}}{2 y-w-r_{1}}\,,
&
\omega_{43} &= \frac{x+y-2 z+\mathbbm{w}\, (y-z)+r_{4}}{x+y-2 z+ \mathbbm{w}\, (y-z)-r_{4}}\,,
\nonumber
\\
\omega_{30} &= \frac{2\, (xy-zw) - (x-z)\, (w-r_1)}{2\, (xy-zw) - (x-z)\, (w+r_1)}\,,\quad
&
\omega_{44} &= \frac{x+y+\mathbbm{w}\, (x+z)+r_{5}}{x+y+\mathbbm{w}\, (x+z)-r_{5}}\,,
\nonumber
\\
\omega_{31} &= \frac{2\, (xy-zw) - (y-z)\, (w-r_1)}{2\, (xy-zw) - (y-z)\, (w+r_1)}\,,
&
\omega_{45} &= \frac{x+y+\mathbbm{w}\, (x-z)+r_{5}}{x+y+\mathbbm{w}\, (x-z)-r_{5}}\,,
\nonumber
\\
\omega_{32} &= \frac{x+y+r_{2}}{x+y-r_{2}}\,,
&
\omega_{46} &= \frac{y-x-\mathbbm{w}\, (x-z)+r_{5}}{y-x-\mathbbm{w}\, (x-z)-r_{5}}\,,
\nonumber
\\
\omega_{33} &= \frac{x-y+r_{2}}{x-y-r_{2}}\,,
&
\omega_{47} &= \frac{x+y-2 z+\mathbbm{w}\, (x-z)+r_{5}}{x+y-2 z+\mathbbm{w}\, (x-z)-r_{5}}\,,
\nonumber
\\
\omega_{34} &= \frac{x+y-2 z+r_{2}}{x+y-2 z-r_{2}}\,,
&
\omega_{48} &= \frac{\xi_{1}+r_{1} r_{3}}{\xi_{1}-r_{1} r_{3}}\,,
\\
\omega_{35} &= \frac{x-y-x y+z w+r_{3}}{x-y-x y+z w-r_{3}}\,,
&
\omega_{49} &= \frac{\xi_{2}+(1+\mathbbm{w})\, r_{2} r_{4}}{\xi_{2}-(1+\mathbbm{w})\, r_{2} r_{4}}\,,
\nonumber
\\
\omega_{36} &= \frac{y-x-x y+z w+r_{3}}{y-x-x y+z w-r_{3}}\,,
&
\omega_{50} &= \frac{\xi_{3}+(1+\mathbbm{w})\, r_{2} r_{5}}{\xi_{3}-(1+\mathbbm{w})\, r_{2} r_{5}}\,,
\nonumber
\\
\omega_{37} &= \frac{x+y+x y-z\, (2+w)+r_{3}}{x+y+x y-z\, (2+w)-r_{3}}\,,
&
\omega_{51} &= \frac{\xi_{4}+(1+\mathbbm{w})\, r_{3} r_{4}}{\xi_{4}-(1+\mathbbm{w})\, r_{3} r_{4}}\,,
\nonumber
\\
\omega_{38} &= \frac{x+y+x y-z\, (2 x-w)+r_{3}}{x+y+x y-z\, (2 x-w)-r_{3}}\,,
&
\omega_{52} &= \frac{\xi_{5}+(1+\mathbbm{w})\, r_{3} r_{5}}{\xi_{5}-(1+\mathbbm{w})\, r_{3} r_{5}}\,,
\nonumber
\\
\omega_{39} &= \frac{x+y+x y-z\, (2 y-w)+r_{3}}{x+y+x y-z\, (2 y-w)-r_{3}}\,,
&
\omega_{53} &= \frac{\xi_{6}+r_{4} r_{5}}{\xi_{6}-r_{4} r_{5}}\,,
\nonumber
\\
\omega_{40} &= \frac{x+y+\mathbbm{w}\, (y+z)+r_{4}}{x+y+\mathbbm{w}\, (y+z)-r_{4}}\,,
\nonumber
\end{align}
\end{itemize}
where the explicit expressions for $\xi_i~ (i = 1, \ldots, 6)$ are presented in Appendix \ref{sec:appA}.

\par
In the $\mathcal{B}_2$ branch, although the alphabet contains five radicals, the canonical MIs exhibit only three distinct radical structures, and therefore, they can be classified into three disjoint subsets, as summarized in Table \ref{tab2}. The MIs in $S_{2\,(00000)}$, as well as those in $S_{2\,(10000)}$ after applying the rationalization transformation $\Phi_1$, can both be expressed in terms of GPLs. In contrast, a GPL representation of the MIs in $S_{2\,(11111)}$ would require the simultaneous rationalization of all five radicals, which is clearly impractical. However, at lower orders of the $\epsilon$-expansion, the radical structure of a canonical MI is relatively simple, and the radical index has not yet reached its maximum value. At $\mathcal{O}(\epsilon)$, the only non-vanishing MI in $S_{2\,(11111)}$ is $g_{52}$, which is independent of all five radicals. The radical indices of the MIs in $S_{2\,(11111)}$ at the second, third, and fourth orders in $\epsilon$ are listed in detail in Table \ref{tab3}.
\begin{table}[htbp]
\footnotesize
\centering
\setlength{\tabcolsep}{6pt}
\renewcommand{\arraystretch}{1.08}
\begin{tabular}{
!{\vrule width 1.2pt}
p{1.6cm}<{\centering}
|
p{3.6cm}<{\centering}
p{4.4cm}<{\centering}
p{1.1cm}<{\centering}
!{\vrule width 1.2pt}
}
\Hline
\textsf{Subset} & \textsf{Radical structure} & \textsf{Master integrals~ ($\boldsymbol{g_i}$)} & \textsf{Count} \\ \Hline
$S_{2\,(00000)}$ & $\{\;\}$ & $1\mbox{-}7, 11\mbox{-}14, 18\mbox{-}25, 40, 42\mbox{-}45$ & 24 \\
$S_{2\,(10000)}$ & $\{r_1\}$ & $8\mbox{-}10, 15\mbox{-}17, 26\mbox{-}39, 41$ & 21 \\
$S_{2\,(11111)}$ & $\{r_1, r_2, r_3, r_4, r_5\}$ & $46\mbox{-}59$ & 14 \\ \Hline
\end{tabular}
\caption{Same as Table \ref{tab1}, but for $\mathcal{B}_2$.}
\label{tab2}
\end{table}
\begin{table}[htbp]
\footnotesize
\centering
\setlength{\tabcolsep}{10pt}
\renewcommand{\arraystretch}{1.05}
\begin{tabular}
{
!{\vrule width 1.2pt}
p{2.1cm}<{\centering}
|
p{3.2cm}<{\centering}
p{2.6cm}<{\centering}
p{2.2cm}<{\centering}
!{\vrule width 1.2pt}
}
\Hline
\multirow{2}{*}{\textsf{Radical index}}
& \multicolumn{3}{c!{\vrule width 1.2pt}}{\textsf{Master integrals~ ($\boldsymbol{g_i}$)}} \\
\cline{2-4}
& $\epsilon^2$ & $\epsilon^3$ & $\epsilon^4$ \\
\Hline
    $(00000)$ & $46,50,51,53,54,59$ & $46$ & \\
    $(10000)$ & $47,48,52,55\mbox{-}58$ & $53,54,59$ & \\
    $(10100)$ & $49$ & $47\mbox{-}49,52,55\mbox{-}58$ & $53,54,59$ \\
    $(10110)$ & & $50$ & \\
    $(10101)$ & & $51$ & \\
    $(10111)$ & & & $47\mbox{-}52,55\mbox{-}58$ \\
    $(11111)$ & & & $46$
\tikz[overlay,remember picture]{
      \begin{pgfonlayer}{background}
      \draw[color=gray!80,line width=1.1pt, dashed,rounded corners=1pt] (-4.6cm,-0.12cm) rectangle (1.11cm,2.9cm);
      \end{pgfonlayer}}\\
\Hline
\end{tabular}
\caption{Radical indices of the MIs in $S_{2\,(11111)}$ up to $\mathcal{O}(\epsilon^4)$.}
\label{tab3}
\end{table}

\par
At $\mathcal{O}(\epsilon^2)$, the MIs in $S_{2\,(11111)}$ involve at most two radicals, $r_1$ and $r_3$. The radical structure of $S_{2\,(11111)}$ at this order closely parallels that of the $\mathcal{B}_{1\,(11)}$ subsystem in the $\mathcal{B}_1$ branch:
\begin{align}
\footnotesize
\setlength{\tabcolsep}{2pt}
\renewcommand{\arraystretch}{1.1}
\begin{tabular}{
  !{\vrule width 1.2pt}
  >{\centering\arraybackslash}p{3.5cm} |
  >{\centering\arraybackslash}p{2.9cm} |
  >{\centering\arraybackslash}p{6.5cm} !{\vrule width 1.2pt}}
\Hline
\textsf{System} & \multicolumn{2}{c!{\vrule width 1.2pt}}{\textsf{Radicals}}
\\[1pt]
\cline{1-3}
$S_{2\,(11111)}$ at $\mathcal{O}(\epsilon^{2})$
& $\displaystyle \sqrt{w\,(4+w)}$ & $\displaystyle \sqrt{\lambda(1+x-z,\, 1+y-z,\, zw-xy)}$
\\[1pt]
\cline{1-3}
$\mathcal{B}_{1\,(11)}$ & $\displaystyle \sqrt{w\,(4+w)}$ & $\displaystyle \sqrt{\lambda(x,\, z,\, w)}$
\\[1pt]
\Hline
\end{tabular}
\nonumber
\end{align}
all radicals are square roots, with radicands given by K\"all\'en-type functions. Motivated by this structural similarity, we construct a variable transformation{\textemdash}analogous to $\Psi${\textemdash}that simultaneously rationalizes $r_1$ and $r_3$:
\begin{equation}
\label{eq:Lambda}
\Lambda(x, z, w) = \Phi_1(w) \,\circ\, \Phi_2(1+x-z, 1+y-z; zw-xy)\,.
\end{equation}
Explicitly, this transformation reads
\begin{equation}
\left\lbrace\,
\begin{aligned}
&
x
=
\frac{(1 + \mathbbm{w})\,y + \mathbbm{w}^2 \big[\, (\mathbbm{x} - \mathbbm{z}) + \mathbbm{x}y\, (1 - \mathbbm{z}) \,\big]}{(1 + \mathbbm{w}) \big[\, 1 + y\,(\mathbbm{x} - \mathbbm{z}) \,\big] + \mathbbm{w}^2\mathbbm{z}\, (1 - \mathbbm{x})}
\\
&
z
=
\frac{(1 + \mathbbm{w})\,(1 + \mathbbm{x}y)\, \big[\, 1 + y\,(1 - \mathbbm{z}) \,\big]}{(1 + \mathbbm{w}) \big[\, 1 + y\,(\mathbbm{x} - \mathbbm{z}) \,\big] + \mathbbm{w}^2\mathbbm{z}\, (1 - \mathbbm{x})}
\\
&
w = \frac{\mathbbm{w}^2}{1+\mathbbm{w}}
\end{aligned}
\right.\,,
\end{equation}
under which $r_1$ takes the same form as in Eq.\eqref{eq:Rr12}, and $r_3$ admits the following rationalized expression in terms of $(\mathbbm{x}, y, \mathbbm{z}, \mathbbm{w})$:
\begin{equation}
r_3
=
\frac{(\mathbbm{w} - y)\, (1 - \mathbbm{x} - \mathbbm{z})\, (y + \mathbbm{w} + y \mathbbm{w})}{(1 + \mathbbm{w}) \big[\, 1 + y\,(\mathbbm{x} - \mathbbm{z}) \,\big] + \mathbbm{w}^2\mathbbm{z}\, (1 - \mathbbm{x})}\,.
\end{equation}
Consequently, including all $\mathcal{O}(\epsilon^2)$ MIs in $S_{2\,(11111)}$, the integrals listed in the first three rows of Table \ref{tab3} can be expressed in terms of GPLs.

\par
As shown in the dashed box of Table \ref{tab3}, at the third and fourth orders in the $\epsilon$-expansion, some MIs in $S_{2\,(11111)}$ exhibit more intricate radical structures with radical indices exceeding $(10100)$, including $\{r_1,r_3,r_4\}$, $\{r_1,r_3,r_5\}$, $\{r_1,r_3,r_4,r_5\}$, and $\{r_1,r_2,r_3,r_4,r_5\}$. At present, however, no transformations are known that can rationalize these more involved radical structures owing to the nested square roots $r_4$ and $r_5$. Nevertheless, by virtue of the first-order recursion relation and the associated second-order relation obtained via integration by parts,
\begin{equation}
\begin{aligned}
&
\vec{g}^{\,(p)}(\vec{x})
=
\int_{\gamma} d\mathbbm{A}\, \vec{g}^{\,(p-1)} + \vec{g}^{\,(p)}(\vec{x}_0)\,,
\\
&
\vec{g}^{\,(p)}(\vec{x})
=
\int_{\gamma} \Big[\, \mathbbm{A}(\vec{x}) - \mathbbm{A} \,\Big]\, d\mathbbm{A}\, \vec{g}^{\,(p-2)}
+
\Big[\, \mathbbm{A}(\vec{x}) - \mathbbm{A}(\vec{x}_0) \,\Big]\, \vec{g}^{\,(p-1)}(\vec{x}_0)
+
\vec{g}^{\,(p)}(\vec{x}_0)\,,
\end{aligned}
\end{equation}
the integrals in the dashed box can be systematically reduced and ultimately expressed as one-fold integrals over GPLs.

\par
For the $\mathcal{B}_2$ branch, the boundary conditions used to fix the integration constants of the MIs are given as follows:
\begin{itemize}
\item permutation symmetries  (28)

Master integrals sharing the same topological structure are related by permutation symmetries of the external kinematic invariants, as exemplified by $g_{18-21}$. Figure \ref{fig4} shows that the $59$ MIs in $\mathcal{B}_2$ reduce to $31$ independent topologies, yielding $28$ homogeneous boundary conditions induced by these permutation symmetries.

\item common topologies with $\mathcal{B}_1$ (16)

Among the $31$ independent topologies in $\mathcal{B}_2$, $16$ coincide with those in $\mathcal{B}_1$, leaving the remaining $15$ fixed by the boundary conditions below.

\item \texttt{HyperInt} evaluation (4)

$g_{22,40,42,44}$ are evaluated analytically using the \texttt{HyperInt} package \cite{Panzer:2014caa}.

\item vanishing conditions (11)

$g_{38,41,46,47,49,50,52,53,55,57,59}$ vanish at $(x,y,z,w)=(0,0,1,0)$.
\end{itemize}

\par
The explicit expressions for all $59$ canonical MIs in $\mathcal{B}_2$ up to $\mathcal{O}(\epsilon^2)$ are collected in Appendix \ref{sec:appC}. The complete analytic expressions up to $\mathcal{O}(\epsilon^4)$ are provided in the supplementary file \texttt{B2.m}.

\section{Analytic extension and numerical check}
\label{sec:6}
\par
As discussed in Section \ref{sec:2}, the topologies associated with the $ZZH$ vertex can be obtained from the four topologies in Fig.\ref{fig1} by replacing the $W$ propagator with a $Z$ propagator. Among these, the two family branches induced by the planar topologies are denoted by $\mathcal{C}_1$ and $\mathcal{C}_2$, respectively. For the $62$ MIs in the $\mathcal{B}_1$ branch associated with the $WWH$ vertex, all integrals remain finite in the limit $m_W^2 \rightarrow m_Z^2$ (i.e. $z \rightarrow -1$), with the exception of $g_{9}$, $g_{13}$, and $g_{51}$. Moreover, the singular parts of these three integrals are not independent, but satisfy the proportionality relation
\begin{equation}
\breve{g}_9 \,:\, \breve{g}_{13} \,:\, \breve{g}_{51} = 1 \,:\, 2 \,:\, -1\,,
\end{equation}
where $\breve{g}$ denotes the singular part of $g$ in the limit $z \rightarrow -1$. This indicates that the $\mathcal{B}_1$ branch contains only a single linearly independent divergence. Accordingly, the canonical MIs for the $\mathcal{B}_1$ branch with the simplest divergent structure can be chosen as
\begin{equation}
E =  E_{r} \cup E_s = \big\{ g_{i \neq 9,13,51},\; 2 g_9 - g_{13},\; g_9+g_{51} \big\} \cup \big\{ g_9 \big\}\,.
\end{equation}
The MIs in the set $E_r$ are regular in the limit $z \rightarrow -1$, whereas $E_s$ contains the only divergent MI. This implies that the $\mathcal{C}_1$ branch associated with the $ZZH$ vertex contains $61$ independent MIs, and a canonical basis can be constructed as
\begin{equation}
\label{eq:canonical-c1}
\vec{h} = \lim_{z \rightarrow -1}\,\left( g_{i\neq 9,13,51},\; 2g_9 - g_{13},\; g_9 + g_{51} \right)\,,
\end{equation}
which has been verified through the IBP reduction using \texttt{Kira}. Specifically, the analytic expressions for $h_{60}$ and $h_{61}$ up to $\mathcal{O}(\epsilon^2)$ are given as follows:
\begin{align}
h_{60} ={}& \lim_{z \rightarrow -1}\,(2g_9 - g_{13})
=
1+\epsilon^{2}\,\pi^2/6 \,,
\nonumber \\
h_{61} ={}& \lim_{z \rightarrow -1}\,(g_{9} + g_{51})
=
\epsilon\, [\, G(a_3;y) + G(a_{22};\mathbbm{x}) \,]
+
\epsilon^2\, \big\{ 4 G(a_3;y)\, [\, G(a_{24};\mathbbm{x}) - G(a_{26};\mathbbm{x}) \,]
\nonumber \\
& - 2G(a_3;\mathbbm{w})\, [\, G(a_{8};y) - G(a_{12};\mathbbm{x}) + G(a_{12};\mathbbm{z}) + G(a_{14};\mathbbm{x}) - G(a_{14};\mathbbm{z}) - G(a_{15};y) \,]
\nonumber \\
& + G(a_{16};\mathbbm{z})\, [\, G(a_2;\mathbbm{x}) - 2 G(a_{12};\mathbbm{x}) - 2 G(a_{14};\mathbbm{x}) - 4 G(a_{24};\mathbbm{x}) + 4 G(a_{26};\mathbbm{x}) \,]
\nonumber \\
& + G(a_1,a_{16};\mathbbm{z}) + 4 G(a_1,a_{22};\mathbbm{x}) + G(a_2,a_{23};\mathbbm{x}) - 4 G(a_3,a_3;y)+ 2 G(a_8,a_3;y)
\nonumber \\
& - 2 G(a_{12},a_{16};\mathbbm{z}) + 2 G(a_{12},a_{22};\mathbbm{x}) - 2 G(a_{12},a_{23};\mathbbm{x}) - 2 G(a_{14},a_{16};\mathbbm{z})
\nonumber \\
& + 2 G(a_{14},a_{22};\mathbbm{x}) - 2 G(a_{14},a_{23};\mathbbm{x}) + 2 G(a_{15},a_3;y) - 4 G(a_{22},a_{22};\mathbbm{x})
\nonumber \\
&- 4 G(a_{24},a_{23};\mathbbm{x}) - 4 G(a_{26},a_{22};\mathbbm{x}) + 4G(a_{26},a_{23};\mathbbm{x})
\big\}\,,
\end{align}
where the weights $a_i$ are defined in Eq.\eqref{eq:B1-weights}, and the variables $\mathbbm{x}$, $\mathbbm{z}$ and $\mathbbm{w}$ satisfy the constraint
\begin{equation}
\mathbbm{z}\,(\mathbbm{x}-1)\,\frac{\mathbbm{w}^2}{1 + \mathbbm{w}}=1\,.
\end{equation}
The analytic expressions up to $\mathcal{O}(\epsilon^4)$ are provided in the supplementary file \texttt{C1.m}. For the $\mathcal{B}_2$ branch associated with the $WWH$ vertex, all MIs are regular in the limit $z \rightarrow -1$. Consequently, the $\mathcal{C}_2$ branch associated with the $ZZH$ vertex contains $59$ independent MIs, and a canonical basis can be conveniently chosen as
\begin{equation}
\vec{h}=(g_{1},\ldots,g_{59})\,|_{z\,=\,-1}\,,
\end{equation}
 which can also be verified using \texttt{Kira}.

\par
To summarize, we derive analytic expressions for the canonical MIs in the four branches induced by the four planar topologies. Most of these MIs can be expressed in terms of GPLs, while a subset of them is represented, at $\mathcal{O}(\epsilon^3)$ and $\mathcal{O}(\epsilon^4)$, by one-fold integrals over GPLs. To check the correctness of our analytic expressions for the MIs, we perform cross-verification against independent numerical results obtained using \texttt{AMFlow} \cite{Liu:2022chg}. The GPLs are evaluated using the \texttt{NumPolyLog}~\cite{Vollinga:2004sn,MMA:MPLG} package, while the one-fold integrals over GPLs are computed with the built-in \texttt{NIntegrate} function in \texttt{Mathematica}. The numerical check is performed at a benchmark point in the Euclidean region,
\begin{equation}
(s, t, u, m_Z^2, m_H^2, m_W^2) = (-11/9,\, -1/15,\, -1/6,\, -19/18,\, -2/5, 1)\,,
\end{equation}
and the results for representative MIs are summarized in Table \ref{tab4}. These integrals are selected to capture the principal nontrivial structures encountered in our calculation: $g_{62}$ exhibits the most intricate radical structure in the $\mathcal{B}_1$ branch; $g_{52}$ serves as a representative MI in the $\mathcal{B}_2$ branch, admitting a one-fold integral representation; and $h_{61}$ is obtained from $g_9 + g_{51}$ in the $\mathcal{B}_1$ branch by taking the limit $z \rightarrow -1$, as shown in Eq.\eqref{eq:canonical-c1}. The numerical results obtained from our analytic expressions exhibit excellent agreement with those computed using the auxiliary mass flow method, with full consistency at the level of $30$ significant digits.
\begin{table}[htbp]
\centering
\footnotesize
\setlength{\tabcolsep}{10pt}
\renewcommand{\arraystretch}{1.05}
  {\setlength{\jot}{-0.6em}
\begin{tabular}{
!{\vrule width 1.2pt}
p{1.2cm}<{\centering}@{\hskip 0.5cm}|
p{1.5cm}<{\centering}@{\hskip 0.4cm}
p{6.8cm}<{\centering}
!{\vrule width 1.2pt}
}
\Hline
\textsf{Integral} & \textsf{Method} & \textsf{Numerical result} \\
\Hline
\multirow{2.6}{*}{$g_{62} \in \mathcal{B}_1$}
                  & \multirow{1.2}{*}{\textsf{Analytic}}        &         $\begin{aligned}
        -\,&0.664957668643736033610081193355\,\epsilon^3 \\
        -\,&4.726392850105022560752360395509\,\epsilon^4
        \end{aligned}$ \\
\cline{2-3}
                  & \multirow{1.2}{*}{\texttt{AMFlow} } & $  \begin{aligned}
                                     -\,&0.664957668643736033610081193355\,\epsilon^3 \\
                                     -\,&4.726392850105022560752360395509\,\epsilon^4
                  \end{aligned} $ \\
\hline
\multirow{4}{*}{$g_{52} \in \mathcal{B}_2$ }
                  & \multirow{1.2}{*}{\textsf{Analytic}}        & $\begin{aligned}
                    &0.064538521137571171672923915684\,\epsilon  \\
                    +\,&0.330242367516912611342417320261\,\epsilon^2 \\
                    +\,&0.708117837983710638396022672956\,\epsilon^3 \\
                    +\,&1.099406021270770791529184043986\,\epsilon^4
                      \end{aligned}$ \\
\cline{2-3}
                  & \multirow{1.2}{*}{\texttt{AMFlow}}        & $\begin{aligned}
                    &0.064538521137571171672923915684\,\epsilon  \\
                    +\,&0.330242367516912611342417320261\,\epsilon^2 \\
                    +\,&0.708117837983710638396022672956\,\epsilon^3 \\
                    +\,&1.099406021270770791529184043986\,\epsilon^4
                      \end{aligned}$ \\
\hline
\multirow{4}{*}{$h_{61} \in \mathcal{C}_1$ }
                  & \multirow{1.2}{*}{\textsf{Analytic} }        & $\begin{aligned}
                     & 0.863046217355342782317657017981\,\epsilon \\
                                    +\,& 7.990926766799493648452542524898\,\epsilon^2 \\
                                    +\,& 28.96621761952415138613880610204\,\epsilon^3 \\
                                    +\,& 122.0034204649560223899898864194\,\epsilon^4
                  \end{aligned} $ \\
\cline{2-3}
  & \multirow{1.2}{*}{\texttt{AMFlow} }     & $\begin{aligned}
                     & 0.863046217355342782317657017981\,\epsilon \\
                                    +\,& 7.990926766799493648452542524898\,\epsilon^2 \\
                                    +\,& 28.96621761952415138613880610204\,\epsilon^3 \\
                                    +\,& 122.0034204649560223899898864194\,\epsilon^4
                  \end{aligned} $ \\
\Hline
\end{tabular}}
\caption{Numerical comparison of the analytic expressions with the corresponding \texttt{AMFlow} results for three representative master integrals.}
\label{tab4}
\end{table}

\section{Summary}
\label{sec:7}
\par
The $gg \rightarrow ZH$ process is an important component of $ZH$ production at the LHC, contributing approximately $10\%$ to the total cross section and becoming increasingly significant in the boosted regime. In this work, we analytically compute the planar MIs for the two-loop NLO EW corrections to $gg \rightarrow ZH$ induced by light-fermion loops. For the two family branches associated with the $WWH$ vertex, $\mathcal{B}_1$ and $\mathcal{B}_2$, we construct canonical bases via the Magnus-expansion method and derive analytic expressions for the MIs up to $\mathcal{O}(\epsilon^4)$ using the canonical differential-equations method. In the $\mathcal{B}_1$ branch, only two square roots appear, and the MIs are classified into four subsets according to their radical structures. For the subsystems induced by these subsets, we apply rationalizing transformations separately, resulting in relatively compact analytic expressions for the MIs in terms of GPLs. In the $\mathcal{B}_2$ branch, five nontrivial radicals are involved that cannot be rationalized simultaneously, and the corresponding MIs are classified into three subsets. For the subset with the most intricate radical structure, certain MIs require their $\epsilon^3$ and $\epsilon^4$ expansions to be expressed as one-fold integrals over GPLs, while the remaining integrals can still be written in terms of GPLs. The canonical master integrals for the two branches associated with the $ZZH$ vertex can be analytically derived from their counterparts in the corresponding $WWH$ vertex branches. To validate our analytic expressions, we numerically evaluate the master integrals at a benchmark point in the Euclidean region and obtain perfect agreement with the results from \texttt{AMFlow}. The analytic expressions for the MIs presented in this work provide essential input for future studies of the complete NLO EW corrections to $gg \rightarrow ZH$.

\vskip 8mm
\noindent{\large\bf Acknowledgments:}
\par
This work is supported by the National Natural Science Foundation of China (Grant No. 12061141005) and the CAS Center for Excellence in Particle Physics (CCEPP).

\appendix
\section{Explicit expressions of $\alpha$, $\beta$ and $\xi$}
\label{sec:appA}
The non-vanishing coefficients $\alpha_{i,j}$ in Eq.\eqref{eq:gB1} are given as follows:
\begin{align}
\alpha_{16,1} &= 1\,,         & \alpha_{32,32} &= -\,4 w\,,     & \alpha_{45,36} &= 2\,(z-x+w)\,, \nonumber \\
\alpha_{16,10} &= 2\,(3+w)\,, & \alpha_{33,7} &= 2\,(1-x)\,,    & \alpha_{45,39} &= z\, (z-x-w)\,, \nonumber \\
\alpha_{16,14} &= 1+w\,,      & \alpha_{33,9} &= 2\,(z-1)\,,    & \alpha_{45,45} &= -\,2 z\,, \nonumber \\
\alpha_{16,16} &= 4\,,        & \alpha_{33,11} &= 1+x\,,        & \alpha_{46,31} &= x+z-w\,,  \\
\alpha_{32,1} &= 1\,,         & \alpha_{33,13} &= -\,1-z\,,     & \alpha_{46,37} &= 2\,(x-z+w)\,, \nonumber \\
\alpha_{32,8} &= 2\,(1-y)\,,  & \alpha_{33,31} &= 2\,(z-x-w)\,, & \alpha_{46,40} &= x\,(x-z-w)\,, \nonumber \\
\alpha_{32,12} &= 1+y\,,      & \alpha_{33,33} &= 4 w\,,        & \alpha_{46,46} &= -\,2 x\,.  \nonumber \\
\alpha_{32,30} &= 2\,(w-y)\,, & \alpha_{45,31} &= x+z-w\,,   \nonumber
\end{align}
The non-vanishing coefficients $\beta_{i,j}$ in Eq.\eqref{eq:gB2} are listed below:
\begin{align}
\beta_{9,1} &= 1\,,           & \qquad \beta_{52,2} &= 2\,(x-1)\,(4 y-w)\,, \nonumber  \\
\beta_{9,4} &= 2\,(3+w)\,,    &  \beta_{52,3} &= 2\,(1-y)\,(4 y-w)\,,  \nonumber  \\
\beta_{9,7} &= 1+w\,,         &  \beta_{52,5} &=(1+x)\,(w-4 y)\,,  \nonumber \\
\beta_{9,9} &= 4\,,           &  \beta_{52,6}  &= (1+y)\, (4 y-w)\,, \nonumber  \\
\beta_{28,1} &= -\,1\,,       &  \beta_{52,17} &= 8 z w\, (w-2 y)\,, \nonumber  \\
\beta_{28,2} &= 2\,(x-1)\,,   &  \beta_{52,26} &= 2\,(x-w)\,(4 y-w)\,, \nonumber \\
\beta_{28,5} &= -\,1-x\,,     &  \beta_{52,27} &= 2\,(w-y)\, (4 y-w)\,, \\
\beta_{28,26} &= 2\,(x-w)\,,  &  \beta_{52,28} &= 8 w\, (2 y-w)\,, \nonumber \\
\beta_{28,28} &= 4 w\,,       &  \beta_{52,29} &= 8 w\, (w-2 y)\,, \nonumber  \\
\beta_{29,1} &= -\,1\,,       &  \beta_{52,49} &= 4 w\,[\,x+y\,(1+x-z)-z\,(y-w)\,]\,, \nonumber  \\
\beta_{29,3} &= 2\,(y-1)\,,   &  \beta_{52,50} &= 8 w\,[\,x+y\,(1-y-z)+w\,(y+z)\,]\,, \nonumber  \\
\beta_{29,6} &= -\,1-y\,,     &  \beta_{52,51} &= 8 w\,[\,x+y\,(1+x-z)\,]\,, \nonumber \\
\beta_{29,27} &= 2\,(y-w)\,,  &  \beta_{52,52} &= 8 z w\,[\,w-y\,(y-w)\,]\,. \nonumber  \\
\beta_{29,29} &= 4 w\,, \nonumber
\end{align}
The explicit expressions for $\xi_i$ in Eq.\eqref{eq:t2-odd} read
\begin{align}
\xi_{1} &= w\, [\,2 + x + y + x y - z\, (4 + w)\,] + 2 x y\,, \nonumber \\
\xi_{2} &= (1+\mathbbm{w})\, [\,(x+y)^2+y \mathbbm{w}\,(x+y)+z \mathbbm{w}\,(x-y)\,] - 2 z \mathbbm{w}^{2}\,(2+\mathbbm{w})\,, \nonumber \\
\xi_{3} &= (1+\mathbbm{w})\, [\,(x+y)^2+x \mathbbm{w}\,(x+y)-z \mathbbm{w}\,(x-y)\,] - 2 z \mathbbm{w}^{2}\,(2+\mathbbm{w})\,, \nonumber \\
\xi_{4} &= (2+w)\,[\,x y\,(1-z)\,(1+w)-z w^{2}\,(2-z)\,] + w z\,[\,x-y\,(1+w)^{2}\,]  \nonumber \\
       &\phantom{=}\, + (1+w)\,[\,x^{2}\,(1+y)+y^{2}\,(1+x)\,(1+w)\,] \,,  \\
\xi_{5} &= (2+w)\,[\,x y\,(1-z)\,(1+w)-z w^{2}\,(2-z)\,] + w z\,[\,y-x\,(1+w)^{2}\,]  \nonumber \\
       &\phantom{=}\, + (1+w)\,[\,y^{2}\,(1+x)+x^{2}\,(1+y)\,(1+w)\,]\,, \nonumber \\
\xi_{6} &= (1+\mathbbm{w})\,[\,(x+y)^{2} + w\,(x y - z w) + z w\, (x+y-z-w-4)\,]\,. \nonumber
\end{align}

\section{Analytic expressions for the MIs of $\mathcal{B}_1$}
\label{sec:appB}
The analytic expressions for the $62$ MIs of the $\mathcal{B}_1$ branch up to $\mathcal{O}(\epsilon^2)$ are listed below:
\begin{align}
g_{1} ={}& 1 +\epsilon^{2}\, \pi^{2} / 2\,, \nonumber \\
g_{2} ={}& 1 -\epsilon\, G(a_{1};x) +\epsilon^{2}\, G(a_{1},a_{1};x)\,, \nonumber \\
g_{3} ={}& 1 -\epsilon\, G(a_{1};y) +\epsilon^{2}\, G(a_{1},a_{1};y)\,, \nonumber \\
g_{4} ={}& 1 -\epsilon\, G(a_{1};z) +\epsilon^{2}\, G(a_{1},a_{1};z)\,, \nonumber \\
g_{5} ={}& 1 -\epsilon\, G(a_{1};w) +\epsilon^{2}\, G(a_{1},a_{1};w)\,, \nonumber \\
g_{6} ={}& -\epsilon\, G(a_{3};\mathbbm{w}) +\epsilon^{2}\, [\,2 G(a_{1},a_{3};\mathbbm{w})+2 G(a_{3},a_{1};\mathbbm{w})-3 G(a_{3},a_{3};\mathbbm{w})+2 G(a_{4},a_{3};\mathbbm{w})\,]\,, \nonumber \\
g_{7} ={}& \epsilon\, G(a_{3};x) +\epsilon^{2}\, [\,G(a_{1},a_{3};x)-4 G(a_{3},a_{3};x)\,]\,, \nonumber \\
g_{8} ={}& \epsilon\, G(a_{3};y) +\epsilon^{2}\, [\,G(a_{1},a_{3};y)-4 G(a_{3},a_{3};y)\,]\,, \nonumber \\
g_{9} ={}& \epsilon\, G(a_{3};z) +\epsilon^{2}\, [\,G(a_{1},a_{3};z)-4 G(a_{3},a_{3};z)\,]\,, \nonumber \\
g_{10} ={}& \epsilon\, G(a_{3};w) +\epsilon^{2}\, [\,G(a_{1},a_{3};w)-4 G(a_{3},a_{3};w)\,]\,, \nonumber \\
g_{11} ={}& -1 +2 \epsilon\, G(a_{3};x) -\epsilon^{2}\, [\,\pi^{2} / 2-4 G(a_{1},a_{3};x)+8 G(a_{3},a_{3};x)\,]\,, \nonumber \\
g_{12} ={}& -1 +2 \epsilon\, G(a_{3};y) -\epsilon^{2}\, [\,\pi^{2} / 2-4 G(a_{1},a_{3};y)+8 G(a_{3},a_{3};y)\,]\,, \nonumber \\
g_{13} ={}& -1 +2 \epsilon\, G(a_{3};z) -\epsilon^{2}\, [\,\pi^{2} / 2-4 G(a_{1},a_{3};z)+8 G(a_{3},a_{3};z)\,]\,, \nonumber \\
g_{14} ={}& -1 +2 \epsilon\, G(a_{3};w) -\epsilon^{2}\, [\,\pi^{2} / 2-4 G(a_{1},a_{3};w)+8 G(a_{3},a_{3};w)\,]\,, \nonumber \\
g_{15} ={}& 0\,, \nonumber \\
g_{16} ={}& 4 \epsilon\, G(a_{3};\mathbbm{w}) +\epsilon^{2}\, \big\{10 G(a_{3},a_{3};\mathbbm{w})-6G(a_{3},a_{5};\mathbbm{w})-6G(a_{3},a_{6};\mathbbm{w})-8 G(a_{4},a_{3};\mathbbm{w})\big\}\,, \nonumber \\
g_{17} ={}& -\epsilon\, G(a_{3};\mathbbm{w}) +\epsilon^{2}\, [\,G(a_{1};y) G(a_{3};\mathbbm{w})-G(a_{3},a_{3};\mathbbm{w})+2 G(a_{4},a_{3};\mathbbm{w})\,]\,, \nonumber \\
g_{18} ={}& -\epsilon\, G(a_{3};\mathbbm{w}) +\epsilon^{2}\, [\,G(a_{1};z) G(a_{3};\mathbbm{w})-G(a_{3},a_{3};\mathbbm{w})+2 G(a_{4},a_{3};\mathbbm{w})\,]\,, \nonumber \\
g_{19} ={}& -\epsilon\, G(a_{3};\mathbbm{w}) +\epsilon^{2}\, [\,G(a_{1};x) G(a_{3};\mathbbm{w})-G(a_{3},a_{3};\mathbbm{w})+2 G(a_{4},a_{3};\mathbbm{w})\,]\,, \nonumber \\
g_{20} ={}& -\epsilon^{2}\, [\,\pi^{2} / 3+G(a_{1};\mathbbm{x}) G(a_{1};\mathbbm{z})-G(a_{2};\mathbbm{x}) G(a_{2};\mathbbm{z})+G(a_{1},a_{2};\mathbbm{x})+G(a_{1},a_{2};\mathbbm{z}) \nonumber \\
&-G(a_{2},a_{1};\mathbbm{x})-G(a_{2},a_{1};\mathbbm{z})\,]\,, \nonumber \\
g_{21} ={}& 0\,, \nonumber \\
g_{22} ={}& 0\,, \nonumber \\
g_{23} ={}& 0\,, \nonumber \\
g_{24} ={}& 0\,, \nonumber \\
g_{25} ={}& 1/2\, \epsilon^{2}\, \big\{G(a_{3};w)\, [\,G(a_{1};y)-G(a_{9};y)\,]-G(a_{1},a_{3};w)-G(a_{3},a_{3};w)+G(a_{9},a_{1};y)\big\}\,, \nonumber \\
g_{26} ={}& 1/2\, \epsilon^{2}\, \big\{[\,G(a_{1};w)-G(a_{3};w)\,]\, [\,G(a_{11};\mathbbm{z})-G(a_{13};\mathbbm{x})\,]-G(a_{1};\mathbbm{x})\, [\,G(a_{3};w)-G(a_{11};\mathbbm{z})\,]\big. \nonumber \\
&+G(a_{2};\mathbbm{z})\, [\,G(a_{3};w)-G(a_{13};\mathbbm{x})\,]-G(a_{11};\mathbbm{z})\, [\,G(a_{2};\mathbbm{x})-G(a_{13};\mathbbm{x})\,]+G(a_{1},a_{11};\mathbbm{z}) \nonumber \\
&+G(a_{1},a_{18};\mathbbm{x})-G(a_{2},a_{11};\mathbbm{z})-G(a_{2},a_{18};\mathbbm{x})-G(a_{3},a_{1};w)+G(a_{3},a_{3};w) \nonumber \\
&\big.+G(a_{11},a_{2};\mathbbm{z})-G(a_{11},a_{11};\mathbbm{z})-G(a_{13},a_{1};\mathbbm{x})+G(a_{13},a_{18};\mathbbm{x})\big\}\,, \nonumber \\
g_{27} ={}& 1/2\, \epsilon^{2}\, \big\{[\,G(a_{1};w)-G(a_{3};w)\,]\, [\,G(a_{11};\mathbbm{x})-G(a_{13};\mathbbm{z})\,]-G(a_{1};\mathbbm{z})\, [\,G(a_{3};w)-G(a_{11};\mathbbm{x})\,]\big. \nonumber \\
&+G(a_{3};w) G(a_{2};\mathbbm{x})+G(a_{1},a_{19};\mathbbm{x})-G(a_{2},a_{19};\mathbbm{x})-G(a_{3},a_{1};w)+G(a_{3},a_{3};w) \nonumber \\
&\big.+G(a_{11},a_{2};\mathbbm{x})-G(a_{11},a_{19};\mathbbm{x})-G(a_{13},a_{1};\mathbbm{z})\big\}\,, \nonumber \\
g_{28} ={}& -2 \epsilon\, G(a_{3};w) +\epsilon^{2}\, \big\{[\,G(a_{1};w)-G(a_{3};w)\,]\, [\,G(a_{11};\mathbbm{z})+G(a_{13};\mathbbm{x})\,]+G(a_{1};\mathbbm{x})\, [\,G(a_{3};w)\big. \nonumber \\
&+G(a_{11};\mathbbm{z})\,]+G(a_{2};\mathbbm{z})\, [\,G(a_{3};w)+G(a_{13};\mathbbm{x})\,]-G(a_{11};\mathbbm{z})\, [\,G(a_{2};\mathbbm{x})+G(a_{13};\mathbbm{x})\,] \nonumber \\
&-G(a_{1},a_{11};\mathbbm{z})+G(a_{1},a_{18};\mathbbm{x})+G(a_{2},a_{11};\mathbbm{z})-G(a_{2},a_{18};\mathbbm{x})+G(a_{3},a_{1};w) \nonumber \\
&\big.+7 G(a_{3},a_{3};w)+G(a_{11},a_{2};\mathbbm{z})-G(a_{11},a_{11};\mathbbm{z})+G(a_{13},a_{1};\mathbbm{x})-G(a_{13},a_{18};\mathbbm{x})\big\}\,, \nonumber \\
g_{29} ={}& -2 \epsilon\, G(a_{19};\mathbbm{x}) +\epsilon^{2}\, \big\{[\,G(a_{1};w)-G(a_{3};w)\,]\, [\,G(a_{11};\mathbbm{x})+G(a_{13};\mathbbm{z})\,]+G(a_{1};\mathbbm{z})\, [\,G(a_{3};w)\big. \nonumber \\
&+G(a_{11};\mathbbm{x})\,]+G(a_{2};\mathbbm{x}) G(a_{3};w)-G(a_{1},a_{19};\mathbbm{x})+G(a_{2},a_{19};\mathbbm{x})+G(a_{3},a_{1};w) \nonumber \\
&\big.-G(a_{3},a_{3};w)+G(a_{11},a_{2};\mathbbm{x})-G(a_{11},a_{19};\mathbbm{x})+G(a_{13},a_{1};\mathbbm{z})+8 G(a_{19},a_{19};\mathbbm{x})\big\}\,, \nonumber \\
g_{30} ={}& -\epsilon^{2}\, \big\{G(a_{3};\mathbbm{w})\, [\,G(a_{8};y)-G(a_{15};y)\,]+G(a_{1},a_{3};y)+2 G(a_{3},a_{3};\mathbbm{w})-G(a_{8},a_{3};y)\big. \nonumber \\
&\big.-G(a_{15},a_{3};y)\big\}\,, \nonumber \\
g_{31} ={}& -\epsilon^{2}\, \big\{G(a_{3};\mathbbm{w})\, [\,G(a_{12};\mathbbm{x})+G(a_{12};\mathbbm{z})-G(a_{14};\mathbbm{x})-G(a_{14};\mathbbm{z})\,]+G(a_{16};\mathbbm{z})\, [\,G(a_{2};\mathbbm{x})\big. \nonumber \\
&-G(a_{12};\mathbbm{x})-G(a_{14};\mathbbm{x})\,]-G(a_{1},a_{16};\mathbbm{z})-G(a_{1},a_{22};\mathbbm{x})+G(a_{2},a_{23};\mathbbm{x})-2 G(a_{3},a_{3};\mathbbm{w}) \nonumber \\
&+G(a_{12},a_{16};\mathbbm{z})+G(a_{12},a_{22};\mathbbm{x})-G(a_{12},a_{23};\mathbbm{x})+G(a_{14},a_{16};\mathbbm{z})+G(a_{14},a_{22};\mathbbm{x}) \nonumber \\
&\big.-G(a_{14},a_{23};\mathbbm{x})\big\}\,, \nonumber \\
g_{32} ={}& 4 \epsilon\, G(a_{3};\mathbbm{w}) -2 \epsilon^{2}\, \big\{G(a_{3};\mathbbm{w})\, [\,G(a_{8};y)+G(a_{15};y)\,]-2 G(a_{3},a_{3};\mathbbm{w})+4 G(a_{4},a_{3};\mathbbm{w})\big. \nonumber \\
&\big.-G(a_{8},a_{3};y)+G(a_{15},a_{3};y)\big\}\,, \nonumber \\
g_{33} ={}& -4 \epsilon\, G(a_{3};\mathbbm{w}) +2 \epsilon^{2}\, \big\{G(a_{3};\mathbbm{w})\, [\,G(a_{12};\mathbbm{x})+G(a_{12};\mathbbm{z})+G(a_{14};\mathbbm{x})+G(a_{14};\mathbbm{z})\,]\big. \nonumber \\
&-G(a_{16};\mathbbm{z})\, [\,G(a_{12};\mathbbm{x})-G(a_{14};\mathbbm{x})\,]-2 G(a_{3},a_{3};\mathbbm{w})+4 G(a_{4},a_{3};\mathbbm{w})+G(a_{12},a_{16};\mathbbm{z}) \nonumber \\
&\big.+G(a_{12},a_{22};\mathbbm{x})-G(a_{12},a_{23};\mathbbm{x})-G(a_{14},a_{16};\mathbbm{z})-G(a_{14},a_{22};\mathbbm{x})+G(a_{14},a_{23};\mathbbm{x})\big\}\,, \nonumber \\
g_{34} ={}& 0\,, \nonumber \\
g_{35} ={}& 0\,, \nonumber \\
g_{36} ={}& 0\,, \nonumber \\
g_{37} ={}& 0\,, \nonumber \\
g_{38} ={}& \epsilon^{2}\, \big\{G(a_{3};\mathbbm{w})\, [\,G(a_{8};y)-G(a_{15};y)\,]+G(a_{1},a_{3};y)+2 G(a_{3},a_{3};\mathbbm{w})-G(a_{8},a_{3};y)\big. \nonumber \\
&\big.-G(a_{15},a_{3};y)\big\}\,, \nonumber \\
g_{39} ={}& \epsilon^{2}\, \big\{G(a_{3};\mathbbm{w})\, [\,G(a_{12};\mathbbm{x})+G(a_{12};\mathbbm{z})-G(a_{14};\mathbbm{x})-G(a_{14};\mathbbm{z})\,]+G(a_{16};\mathbbm{z})\, [\,G(a_{2};\mathbbm{x})\big. \nonumber \\
&-G(a_{12};\mathbbm{x})-G(a_{14};\mathbbm{x})\,]-G(a_{1},a_{16};\mathbbm{z})-G(a_{1},a_{22};\mathbbm{x})+G(a_{2},a_{23};\mathbbm{x})-2 G(a_{3},a_{3};\mathbbm{w}) \nonumber \\
&+G(a_{12},a_{16};\mathbbm{z})+G(a_{12},a_{22};\mathbbm{x})-G(a_{12},a_{23};\mathbbm{x})+G(a_{14},a_{16};\mathbbm{z})+G(a_{14},a_{22};\mathbbm{x}) \nonumber \\
&\big.-G(a_{14},a_{23};\mathbbm{x})\big\}\,, \nonumber \\
g_{40} ={}& \epsilon^{2}\, \big\{G(a_{3};\mathbbm{w})\, [\,G(a_{12};\mathbbm{x})+G(a_{12};\mathbbm{z})-G(a_{14};\mathbbm{x})-G(a_{14};\mathbbm{z})\,]+G(a_{16};\mathbbm{z})\, [\,G(a_{2};\mathbbm{x})\big. \nonumber \\
&-G(a_{12};\mathbbm{x})-G(a_{14};\mathbbm{x})\,]-G(a_{1},a_{16};\mathbbm{z})-G(a_{1},a_{22};\mathbbm{x})+G(a_{2},a_{23};\mathbbm{x})-2 G(a_{3},a_{3};\mathbbm{w}) \nonumber \\
&+G(a_{12},a_{16};\mathbbm{z})+G(a_{12},a_{22};\mathbbm{x})-G(a_{12},a_{23};\mathbbm{x})+G(a_{14},a_{16};\mathbbm{z})+G(a_{14},a_{22};\mathbbm{x}) \nonumber \\
&\big.-G(a_{14},a_{23};\mathbbm{x})\big\}\,, \nonumber \\
g_{41} ={}& 0\,, \nonumber \\
g_{42} ={}& 0\,, \nonumber \\
g_{43} ={}& 0\,, \nonumber \\
g_{44} ={}& 0\,, \nonumber \\
g_{45} ={}& 0\,, \nonumber \\
g_{46} ={}& 0\,, \nonumber \\
g_{47} ={}& 0\,, \nonumber \\
g_{48} ={}& 1 -\epsilon\, [\,2 G(a_{1};y)-G(a_{1};w)+G(a_{1};\mathbbm{x})-G(a_{1};\mathbbm{z})-G(a_{2};\mathbbm{x})+G(a_{2};\mathbbm{z})\,]+\epsilon^{2}\, \big\{2 \pi^{2} / 3\big. \nonumber \\
&-2 G(a_{1};y)\, [\,G(a_{1};w)-G(a_{1};\mathbbm{x})+G(a_{1};\mathbbm{z})-G(a_{2};\mathbbm{z})+G(a_{21};\mathbbm{x})\,] \nonumber \\
&-G(a_{1};w)\, [\,G(a_{1};\mathbbm{x})-G(a_{1};\mathbbm{z})+G(a_{2};\mathbbm{x})+G(a_{2};\mathbbm{z})-2 G(a_{7};y)-2 G(a_{10};y) \nonumber \\
&-2 G(a_{21};\mathbbm{x})\,]-G(a_{1};\mathbbm{x})\, [\,G(a_{1};\mathbbm{z})-G(a_{2};\mathbbm{z})\,]-G(a_{1};\mathbbm{z})\, [\,G(a_{2};\mathbbm{x})-2 G(a_{10};y) \nonumber \\
&-2 G(a_{21};\mathbbm{x})\,]-G(a_{2};\mathbbm{x}) G(a_{2};\mathbbm{z})+4 G(a_{1},a_{1};y)+G(a_{1},a_{1};w)+G(a_{1},a_{1};\mathbbm{x}) \nonumber \\
&+G(a_{1},a_{1};\mathbbm{z})-G(a_{1},a_{2};\mathbbm{x})-G(a_{1},a_{2};\mathbbm{z})-G(a_{2},a_{1};\mathbbm{x})-G(a_{2},a_{1};\mathbbm{z}) \nonumber \\
&\big.-G(a_{2},a_{2};\mathbbm{x})+G(a_{2},a_{2};\mathbbm{z})-2 G(a_{7},a_{1};y)-2 G(a_{10},a_{1};y)+2 G(a_{21},a_{2};\mathbbm{x})\big\}\,, \nonumber \\
g_{49} ={}& -\epsilon\, G(a_{3};\mathbbm{w}) +\epsilon^{2}\, \big\{G(a_{3};\mathbbm{w})\, [\,2 G(a_{1};y)+G(a_{1};\mathbbm{x})-G(a_{1};\mathbbm{z})-G(a_{2};\mathbbm{x})+G(a_{2};\mathbbm{z})\,]\big. \nonumber \\
&\big.-2 G(a_{1},a_{3};\mathbbm{w})-2 G(a_{3},a_{1};\mathbbm{w})+G(a_{3},a_{3};\mathbbm{w})+2 G(a_{4},a_{3};\mathbbm{w})\big\}\,, \nonumber \\
g_{50} ={}& 0\,, \nonumber \\
g_{51} ={}& \epsilon\, [\,G(a_{3};y)-G(a_{16};\mathbbm{z})+G(a_{22};\mathbbm{x})-G(a_{23};\mathbbm{x})\,]+2 \epsilon^{2}\, \big\{2 G(a_{3};y)\, [\,G(a_{24};\mathbbm{x}) \nonumber \\
&-G(a_{26};\mathbbm{x})\,]-G(a_{3};\mathbbm{w})\, [\,G(a_{8};y)-G(a_{12};\mathbbm{x})+G(a_{12};\mathbbm{z})+G(a_{14};\mathbbm{x})-G(a_{14};\mathbbm{z}) \nonumber \\
&-G(a_{15};y)\,]-G(a_{16};\mathbbm{z})\, [\,G(a_{12};\mathbbm{x})+G(a_{14};\mathbbm{x})-2 G(a_{23};\mathbbm{x})+2 G(a_{24};\mathbbm{x}) \nonumber \\
&-2 G(a_{26};\mathbbm{x})\,]+2 G(a_{1},a_{22};\mathbbm{x})-2 G(a_{3},a_{3};y)+G(a_{8},a_{3};y)-G(a_{12},a_{16};\mathbbm{z}) \nonumber \\
&+G(a_{12},a_{22};\mathbbm{x})-G(a_{12},a_{23};\mathbbm{x})-G(a_{14},a_{16};\mathbbm{z})+G(a_{14},a_{22};\mathbbm{x})-G(a_{14},a_{23};\mathbbm{x}) \nonumber \\
&+G(a_{15},a_{3};y)+2 G(a_{16},a_{16};\mathbbm{z})-2 G(a_{22},a_{22};\mathbbm{x})+2 G(a_{23},a_{23};\mathbbm{x})-2 G(a_{24},a_{23};\mathbbm{x}) \nonumber \\
&\big.-2 G(a_{26},a_{22};\mathbbm{x})+2 G(a_{26},a_{23};\mathbbm{x})\big\}\,, \nonumber \\
g_{52} ={}& -1 +\epsilon\, [\,G(a_{1};w)+G(a_{1};\mathbbm{x})+G(a_{2};\mathbbm{z})+G(a_{3};y)\,]-\epsilon^{2}\, \big\{G(a_{1};w)\, [\,G(a_{1};\mathbbm{x}) \nonumber \\
&+G(a_{2};\mathbbm{z})+G(a_{3};y)\,]+G(a_{1};\mathbbm{x})\, [\,G(a_{2};\mathbbm{z})+G(a_{3};y)\,]+G(a_{3};y)\, [\,G(a_{2};\mathbbm{z}) \nonumber \\
&-G(a_{3};w)-G(a_{11};\mathbbm{z})-G(a_{21};\mathbbm{x})\,]+G(a_{11};\mathbbm{z})\, [\,G(a_{10};y)+G(a_{21};\mathbbm{x})\,] \nonumber \\
&+G(a_{3};w) G(a_{7};y)+G(a_{1},a_{1};w)+G(a_{1},a_{1};\mathbbm{x})-G(a_{1},a_{3};y)+G(a_{1},a_{3};w) \nonumber \\
&+G(a_{1},a_{11};\mathbbm{z})+G(a_{2},a_{2};\mathbbm{z})+4 G(a_{3},a_{3};y)-G(a_{7},a_{3};y)-G(a_{10},a_{3};y) \nonumber \\
&\big.+G(a_{21},a_{18};\mathbbm{x})\big\}\,, \nonumber \\
g_{53} ={}& \epsilon\, [\,G(a_{1};y)-G(a_{1};w)-G(a_{1};\mathbbm{z})-G(a_{2};\mathbbm{x})\,]+\epsilon^{2}\, \big\{G(a_{1};y)\, [\,G(a_{21};\mathbbm{x})-G(a_{25};\mathbbm{x})\,] \nonumber \\
&+G(a_{1};w)\, [\,G(a_{1};\mathbbm{z})+G(a_{2};\mathbbm{x})-G(a_{21};\mathbbm{x})+G(a_{25};\mathbbm{x})\,]+G(a_{1};\mathbbm{z})\, [\,G(a_{2};\mathbbm{x}) \nonumber \\
&-G(a_{21};\mathbbm{x})+G(a_{25};\mathbbm{x})\,]-G(a_{1},a_{1};y)+G(a_{1},a_{1};w)+G(a_{1},a_{1};\mathbbm{z})+G(a_{1},a_{19};\mathbbm{x}) \nonumber \\
&+G(a_{2},a_{2};\mathbbm{x})+G(a_{2},a_{19};\mathbbm{x})\big.-G(a_{21},a_{2};\mathbbm{x})+G(a_{25},a_{2};\mathbbm{x})-G(a_{25},a_{19};\mathbbm{x})\big\}\,, \nonumber \\
g_{54} ={}& \epsilon\, G(a_{3};y) -\epsilon^{2}\, \big\{G(a_{3};y)\, [\,G(a_{1};w)+G(a_{1};\mathbbm{x})+G(a_{2};\mathbbm{z})-G(a_{3};w)-G(a_{11};\mathbbm{z}) \nonumber \\
&-G(a_{21};\mathbbm{x})\,]-G(a_{11};\mathbbm{z})\, [\,G(a_{2};\mathbbm{x})-G(a_{10};y)-G(a_{21};\mathbbm{x})\,]+G(a_{3};w) G(a_{7};y) \nonumber \\
&+G(a_{1},a_{3};y)-G(a_{2},a_{18};\mathbbm{x})+4 G(a_{3},a_{3};y)-G(a_{7},a_{3};y)-G(a_{10},a_{3};y) \nonumber \\
&\big.+G(a_{21},a_{18};\mathbbm{x})\big\}\,, \nonumber \\
g_{55} ={}& \epsilon^{2}\, \big\{[\,G(a_{1};y)-G(a_{1};w)-G(a_{1};\mathbbm{z})\,]\, [\,G(a_{21};\mathbbm{x})-G(a_{25};\mathbbm{x})\,]+G(a_{2},a_{19};\mathbbm{x}) \nonumber \\
&\big.-G(a_{21},a_{2};\mathbbm{x})+G(a_{25},a_{2};\mathbbm{x})-G(a_{25},a_{19};\mathbbm{x})\big\}\,, \nonumber \\
g_{56} ={}& -\epsilon^{2}\, \big\{G(a_{3};\mathbbm{w})\, [\,G(a_{8};y)-G(a_{15};y)\,]+G(a_{1},a_{3};y)+2 G(a_{3},a_{3};\mathbbm{w})\big. \nonumber \\
&\big.-G(a_{8},a_{3};y)-G(a_{15},a_{3};y)\big\}\,, \nonumber \\
g_{57} ={}& 0\,, \nonumber \\
g_{58} ={}& -2 \epsilon\, G(a_{3};y) +\epsilon^{2}\, \big\{2 G(a_{3};y)\, [\,G(a_{1};\mathbbm{x})+2 G(a_{1};\mathbbm{w})+G(a_{2};\mathbbm{z})-G(a_{5};\mathbbm{w})-G(a_{6};\mathbbm{w}) \nonumber \\
&-G(a_{16};\mathbbm{z})-G(a_{26};\mathbbm{x})\,]-G(a_{3};\mathbbm{w})\, [\,2 G(a_{8};y)-G(a_{12};\mathbbm{x})+G(a_{12};\mathbbm{z})+G(a_{14};\mathbbm{x}) \nonumber \\
&-G(a_{14};\mathbbm{z})-2 G(a_{15};y)+2 G(a_{17};y)\,]-G(a_{16};\mathbbm{z})\, [\,G(a_{12};\mathbbm{x})+G(a_{14};\mathbbm{x})-2 G(a_{20};y) \nonumber \\
&-2 G(a_{26};\mathbbm{x})\,]+2 G(a_{17};y)\, [\,G(a_{5};\mathbbm{w})+G(a_{6};\mathbbm{w})\,]-2 G(a_{1},a_{3};y)-2 G(a_{1},a_{3};\mathbbm{w}) \nonumber \\
&+2 G(a_{1},a_{5};\mathbbm{w})+2 G(a_{1},a_{6};\mathbbm{w})+2 G(a_{1},a_{16};\mathbbm{z})+8 G(a_{3},a_{3};y)-G(a_{3},a_{3};\mathbbm{w}) \nonumber \\
&-G(a_{3},a_{5};\mathbbm{w})-G(a_{3},a_{6};\mathbbm{w})+2 G(a_{8},a_{3};y)-G(a_{12},a_{16};\mathbbm{z})+G(a_{12},a_{22};\mathbbm{x}) \nonumber \\
&-G(a_{12},a_{23};\mathbbm{x})-G(a_{14},a_{16};\mathbbm{z})+G(a_{14},a_{22};\mathbbm{x})-G(a_{14},a_{23};\mathbbm{x})+2 G(a_{15},a_{3};y) \nonumber \\
&\big.-2 G(a_{17},a_{3};y)-2 G(a_{20},a_{3};y)-2 G(a_{26},a_{22};\mathbbm{x})+2 G(a_{26},a_{23};\mathbbm{x})\big\}\,, \nonumber \\
g_{59} ={}& -\epsilon^{2}\, \big\{2 [\,G(a_{1};y)-G(a_{1};\mathbbm{z})-2 G(a_{1};\mathbbm{w})\,]\, [\,G(a_{24};\mathbbm{x})-G(a_{27};\mathbbm{x})\,] \nonumber \\
&-2 G(a_{3};y)\, [\,G(a_{24};\mathbbm{x})-G(a_{26};\mathbbm{x})\,]+G(a_{3};\mathbbm{w})\, [\,G(a_{8};y)-G(a_{12};\mathbbm{x})+G(a_{12};\mathbbm{z}) \nonumber \\
&+G(a_{14};\mathbbm{x})-G(a_{14};\mathbbm{z})-G(a_{15};y)+2 G(a_{24};\mathbbm{x})-2 G(a_{27};\mathbbm{x})\,]-G(a_{16};\mathbbm{z})\, [\,2 G(a_{2};\mathbbm{x}) \nonumber \\
&-G(a_{12};\mathbbm{x})-G(a_{14};\mathbbm{x})-2 G(a_{24};\mathbbm{x})+2 G(a_{26};\mathbbm{x})\,]+2 G(a_{1},a_{3};y)-2 G(a_{1},a_{16};\mathbbm{z}) \nonumber \\
&+2 G(a_{2},a_{22};\mathbbm{x})-2 G(a_{2},a_{23};\mathbbm{x})-G(a_{8},a_{3};y)+G(a_{12},a_{16};\mathbbm{z})-G(a_{12},a_{22};\mathbbm{x}) \nonumber \\
&+G(a_{12},a_{23};\mathbbm{x})+G(a_{14},a_{16};\mathbbm{z})-G(a_{14},a_{22};\mathbbm{x})+G(a_{14},a_{23};\mathbbm{x})-G(a_{15},a_{3};y) \nonumber \\
&-2 G(a_{24},a_{2};\mathbbm{x})+2 G(a_{24},a_{23};\mathbbm{x})+2 G(a_{26},a_{22};\mathbbm{x})-2 G(a_{26},a_{23};\mathbbm{x})+2 G(a_{27},a_{2};\mathbbm{x}) \nonumber \\
&\big.-2 G(a_{27},a_{22};\mathbbm{x})\big\}\,, \nonumber \\
g_{60} ={}& -2 \epsilon\, G(a_{3};\mathbbm{w}) +\epsilon^{2}\, \big\{G(a_{3};\mathbbm{w})\, [\,2 G(a_{1};\mathbbm{x})+2 G(a_{2};\mathbbm{z})+2 G(a_{8};y) \nonumber \\
&-G(a_{12};\mathbbm{x})-G(a_{12};\mathbbm{z})-G(a_{14};\mathbbm{x})-G(a_{14};\mathbbm{z})+2 G(a_{15};y)\,]+G(a_{16};\mathbbm{z})\, [\,G(a_{12};\mathbbm{x}) \nonumber \\
&-G(a_{14};\mathbbm{x})\,]+4 G(a_{1},a_{3};\mathbbm{w})+4 G(a_{3},a_{1};\mathbbm{w})-3 G(a_{3},a_{3};\mathbbm{w})-3 G(a_{3},a_{5};\mathbbm{w}) \nonumber \\
&-3 G(a_{3},a_{6};\mathbbm{w})+4 G(a_{4},a_{3};\mathbbm{w})-2 G(a_{8},a_{3};y)-G(a_{12},a_{16};\mathbbm{z})-G(a_{12},a_{22};\mathbbm{x}) \nonumber \\
&\big.+G(a_{12},a_{23};\mathbbm{x})+G(a_{14},a_{16};\mathbbm{z})+G(a_{14},a_{22};\mathbbm{x})-G(a_{14},a_{23};\mathbbm{x})+2 G(a_{15},a_{3};y)\big\}\,, \nonumber \\
g_{61} ={}& \epsilon^{2}\, \big\{G(a_{3};\mathbbm{w})\, [\,2 G(a_{1};y)-2 G(a_{1};\mathbbm{z})-2 G(a_{2};\mathbbm{x})-G(a_{8};y)+G(a_{12};\mathbbm{x}) \nonumber \\
&+G(a_{12};\mathbbm{z})+G(a_{14};\mathbbm{x})+G(a_{14};\mathbbm{z})-G(a_{15};y)\,]-G(a_{16};\mathbbm{z})\, [\,G(a_{12};\mathbbm{x}) \nonumber \\
&-G(a_{14};\mathbbm{x})\,]-4 G(a_{1},a_{3};\mathbbm{w})-4 G(a_{3},a_{1};\mathbbm{w})+4 G(a_{3},a_{3};\mathbbm{w})+G(a_{8},a_{3};y) \nonumber \\
&+G(a_{12},a_{16};\mathbbm{z})+G(a_{12},a_{22};\mathbbm{x})-G(a_{12},a_{23};\mathbbm{x})-G(a_{14},a_{16};\mathbbm{z})-G(a_{14},a_{22};\mathbbm{x}) \nonumber \\
&\big.+G(a_{14},a_{23};\mathbbm{x})-G(a_{15},a_{3};y)\big\}\,, \nonumber \\
g_{62} ={}& 0\,,
\end{align}
where the weights of the GPLs read
\begin{align}
\label{eq:B1-weights}
a_{1} &= 0\,, & a_{2} &= 1\,, & a_{3} &= -\, 1\,, \nonumber \\
a_{4} &= -\, 2\,, & a_{5} &= (-1)^{2/3}\,, & a_{6} &= -\, (-1)^{1/3}\,, \nonumber \\
a_{7} &= w\,, & a_{8} &= \mathbbm{w}\,, & a_{9} &= 1+w\,, \nonumber \\
a_{10} &= w \mathbbm{z}\,, & a_{11} &= -\, \frac{1}{w}\,, & a_{12} &= -\, \frac{1}{\mathbbm{w}}\,, \nonumber \\
a_{13} &= 1+\frac{1}{w}\,, & a_{14} &= 1+\frac{1}{\mathbbm{w}}\,, & a_{15} &= -\, \frac{\mathbbm{w}}{1+\mathbbm{w}}\,, \\
a_{16} &= -\, \frac{1+\mathbbm{w}}{\mathbbm{w}^{2}}\,, & a_{17} &= \frac{\mathbbm{w}^{2}}{1+\mathbbm{w}}\,, & a_{18} &= 1+\frac{1}{w \mathbbm{z}}\,, \nonumber \\
a_{19} &= \frac{1}{w\, (\mathbbm{z}-1)}\,, & a_{20} &= \frac{\mathbbm{z} \mathbbm{w}^{2}}{1+\mathbbm{w}}\,, & a_{21} &= 1-\frac{y}{w \mathbbm{z}}\,, \nonumber \\
a_{22} &= \frac{1+\mathbbm{w}}{\mathbbm{w}^{2}\, (\mathbbm{z}-1)}\,, & a_{23} &= 1 + \frac{1+\mathbbm{w}}{\mathbbm{z} \mathbbm{w}^{2}}\,, & a_{24} &= 1-\frac{y\, (1+\mathbbm{w})}{\mathbbm{z} \mathbbm{w}^{2}}\,, \nonumber \\
a_{25} &= \frac{w \mathbbm{z}-y}{w\, (y+\mathbbm{z}-y \mathbbm{z})}\,, & a_{26} &= \frac{\mathbbm{z} \mathbbm{w}^{2}-y\, (1+\mathbbm{w})}{\mathbbm{w}^{2} \big[\,1+y\, (1-\mathbbm{z}) \big]}\,, & a_{27} &= \frac{\mathbbm{z} \mathbbm{w}^{2}-y\, (1+\mathbbm{w})}{\mathbbm{w}^{2} \big[\, \mathbbm{z}+y\, (1-\mathbbm{z})\,\big]}\,. \nonumber
\end{align}

\section{Analytic expressions for the MIs of $\mathcal{B}_2$}
\label{sec:appC}
Below, we present the analytic expressions for the $59$ MIs of the $\mathcal{B}_2$ branch, up to $\mathcal{O}(\epsilon^2)$:
\begin{align}
g_{1} ={}& 1 +\epsilon^{2}\, \pi^{2} / 2\,, \nonumber \\
g_{2} ={}& \epsilon\, G(a_{3};x) +\epsilon^{2}\, [\,G(a_{1},a_{3};x)-4 G(a_{3},a_{3};x)\,]\,, \nonumber \\
g_{3} ={}& \epsilon\, G(a_{3};y) +\epsilon^{2}\, [\,G(a_{1},a_{3};y)-4 G(a_{3},a_{3};y)\,]\,, \nonumber \\
g_{4} ={}& \epsilon\, G(a_{3};w) +\epsilon^{2}\, [\,G(a_{1},a_{3};w)-4 G(a_{3},a_{3};w)\,]\,, \nonumber \\
g_{5} ={}& -1 +2 \epsilon\, G(a_{3};x) -\epsilon^{2}\, [\,\pi^{2} / 2-4 G(a_{1},a_{3};x)+8 G(a_{3},a_{3};x)\,]\,, \nonumber \\
g_{6} ={}& -1 +2 \epsilon\, G(a_{3};y) -\epsilon^{2}\, [\,\pi^{2} / 2-4 G(a_{1},a_{3};y)+8 G(a_{3},a_{3};y)\,]\,, \nonumber \\
g_{7} ={}& -1 +2 \epsilon\, G(a_{3};w) -\epsilon^{2}\, [\,\pi^{2} / 2-4 G(a_{1},a_{3};w)+8 G(a_{3},a_{3};w)\,]\,, \nonumber \\
g_{8} ={}& 0\,, \nonumber \\
g_{9} ={}& 4 \epsilon\, G(a_{3};\mathbbm{w}) +\epsilon^{2}\, \big\{10 G(a_{3},a_{3};\mathbbm{w})-6G(a_{3},a_{5};\mathbbm{w})-6G(a_{3},a_{6};\mathbbm{w})\big. \nonumber \\
&\big.-\,8 G(a_{4},a_{3};\mathbbm{w})\big\}\,, \nonumber \\
g_{10} ={}& -\epsilon\, G(a_{3};\mathbbm{w}) +\epsilon^{2}\, [\,2 G(a_{1},a_{3};\mathbbm{w})+2 G(a_{3},a_{1};\mathbbm{w})-3 G(a_{3},a_{3};\mathbbm{w}) \nonumber \\
&+2 G(a_{4},a_{3};\mathbbm{w})\,]\,, \nonumber \\
g_{11} ={}& 1 -\epsilon\, G(a_{1};x) +\epsilon^{2}\, G(a_{1},a_{1};x)\,, \nonumber \\
g_{12} ={}& 1 -\epsilon\, G(a_{1};y) +\epsilon^{2}\, G(a_{1},a_{1};y)\,, \nonumber \\
g_{13} ={}& 1 -\epsilon\, G(a_{1};z) +\epsilon^{2}\, G(a_{1},a_{1};z)\,, \nonumber \\
g_{14} ={}& 1 -\epsilon\, G(a_{1};w) +\epsilon^{2}\, G(a_{1},a_{1};w)\,, \nonumber \\
g_{15} ={}& -\epsilon\, G(a_{3};\mathbbm{w}) +\epsilon^{2}\, [\,G(a_{1};x) G(a_{3};\mathbbm{w})-G(a_{3},a_{3};\mathbbm{w})+2 G(a_{4},a_{3};\mathbbm{w})\,]\,, \nonumber \\
g_{16} ={}& -\epsilon\, G(a_{3};\mathbbm{w}) +\epsilon^{2}\, [\,G(a_{1};y) G(a_{3};\mathbbm{w})-G(a_{3},a_{3};\mathbbm{w})+2 G(a_{4},a_{3};\mathbbm{w})\,]\,, \nonumber \\
g_{17} ={}& -\epsilon\, G(a_{3};\mathbbm{w}) +\epsilon^{2}\, [\,G(a_{1};z) G(a_{3};\mathbbm{w})-G(a_{3},a_{3};\mathbbm{w})+2 G(a_{4},a_{3};\mathbbm{w})\,]\,, \nonumber \\
g_{18} ={}& 0\,, \nonumber \\
g_{19} ={}& 0\,, \nonumber \\
g_{20} ={}& 0\,, \nonumber \\
g_{21} ={}& 0\,, \nonumber \\
g_{22} ={}& \epsilon^{2}\, \big\{G(a_{3};w)\, [\,G(a_{1};x)-G(a_{11};x)\,]-G(a_{1},a_{3};w)-G(a_{3},a_{3};w)+G(a_{11},a_{1};x)\big\}\,, \nonumber \\
g_{23} ={}& \epsilon^{2}\, \big\{G(a_{3};w)\, [\,G(a_{1};y)-G(a_{11};y)\,]-G(a_{1},a_{3};w)-G(a_{3},a_{3};w)+G(a_{11},a_{1};y)\big\}\,, \nonumber \\
g_{24} ={}& -\epsilon^{2}\, [\,G(a_{1};z) G(a_{10};x)-G(a_{1},a_{3};x)+G(a_{2},a_{1};z)-G(a_{10},a_{3};x)\,]\,, \nonumber \\
g_{25} ={}& -\epsilon^{2}\, [\,G(a_{1};z) G(a_{10};y)-G(a_{1},a_{3};y)+G(a_{2},a_{1};z)-G(a_{10},a_{3};y)\,]\,, \nonumber \\
g_{26} ={}& -\epsilon^{2}\, \big\{G(a_{3};\mathbbm{w})\, [\,G(a_{9};x)-G(a_{18};x)\,]+G(a_{1},a_{3};x)+2 G(a_{3},a_{3};\mathbbm{w})-G(a_{9},a_{3};x)\big. \nonumber \\
&\big.-\,G(a_{18},a_{3};x)\big\}\,, \nonumber \\
g_{27} ={}& -\epsilon^{2}\, \big\{G(a_{3};\mathbbm{w})\, [\,G(a_{9};y)-G(a_{18};y)\,]+G(a_{1},a_{3};y)+2 G(a_{3},a_{3};\mathbbm{w})-G(a_{9},a_{3};y)\big. \nonumber \\
&\big.-\,G(a_{18},a_{3};y)\big\}\,, \nonumber \\
g_{28} ={}& -4 \epsilon\, G(a_{3};\mathbbm{w}) +2 \epsilon^{2}\, \big\{G(a_{3};\mathbbm{w})\, [\,G(a_{9};x)+G(a_{18};x)\,]-2 G(a_{3},a_{3};\mathbbm{w})\big. \nonumber \\
&\big.+\,4 G(a_{4},a_{3};\mathbbm{w})-G(a_{9},a_{3};x)+G(a_{18},a_{3};x)\big\}\,, \nonumber \\
g_{29} ={}& -4 \epsilon\, G(a_{3};\mathbbm{w}) +2 \epsilon^{2}\, \big\{G(a_{3};\mathbbm{w})\, [\,G(a_{9};y)+G(a_{18};y)\,]-2 G(a_{3},a_{3};\mathbbm{w})\big. \nonumber \\
&\big.+\,4 G(a_{4},a_{3};\mathbbm{w})-G(a_{9},a_{3};y)+G(a_{18},a_{3};y)\big\}\,, \nonumber \\
g_{30} ={}& 0\,, \nonumber \\
g_{31} ={}& 0\,, \nonumber \\
g_{32} ={}& 0\,, \nonumber \\
g_{33} ={}& 0\,, \nonumber \\
g_{34} ={}& 0\,, \nonumber \\
g_{35} ={}& 0\,, \nonumber \\
g_{36} ={}& \epsilon^{2}\, \big\{G(a_{3};\mathbbm{w})\, [\,G(a_{9};x)-G(a_{18};x)\,]+G(a_{1},a_{3};x)+2 G(a_{3},a_{3};\mathbbm{w})-G(a_{9},a_{3};x)\big. \nonumber \\
&\big.-\,G(a_{18},a_{3};x)\big\}\,, \nonumber \\
g_{37} ={}& \epsilon^{2}\, \big\{G(a_{3};\mathbbm{w})\, [\,G(a_{9};y)-G(a_{18};y)\,]+G(a_{1},a_{3};y)+2 G(a_{3},a_{3};\mathbbm{w})-G(a_{9},a_{3};y)\big. \nonumber \\
&\big.-\,G(a_{18},a_{3};y)\big\}\,, \nonumber \\
g_{38} ={}& 0\,, \nonumber \\
g_{39} ={}& 0\,, \nonumber \\
g_{40} ={}& -2 \epsilon\, [\,G(a_{1};x)+G(a_{1};y)-G(a_{1};z)-G(a_{1};w)\,] +\epsilon^{2}\, \big\{\pi^{2}-2 G(a_{1};z)\, [\,G(a_{1};w) \nonumber \\
&-G(a_{7};x)-G(a_{7};y)+G(a_{17};x)\,]+2 G(a_{1};w)\, [\,G(a_{8};x)+G(a_{8};y)-G(a_{17};x)\,] \nonumber \\
&+2 G(a_{1};y)\, G(a_{17};x)+2 G(a_{1},a_{1};x)+2 G(a_{1},a_{1};y)-2 G(a_{7},a_{1};x)-2 G(a_{7},a_{1};y) \nonumber \\
&\big.-\,2 G(a_{8},a_{1};x)-2 G(a_{8},a_{1};y)+2 G(a_{17},a_{1};x)\big\}\,, \nonumber \\
g_{41} ={}& 2 \epsilon^{2}\, \big\{G(a_{3};\mathbbm{w})\, [\,G(a_{1};x)+G(a_{1};y)-G(a_{1};z)\,]-2G(a_{1},a_{3};\mathbbm{w})-2G(a_{3},a_{1};\mathbbm{w}) \nonumber \\
 &+2G(a_{3},a_{3};\mathbbm{w})\big\}\,, \nonumber \\
g_{42} ={}& \epsilon\, [\,G(a_{1};x)-G(a_{1};z)\,] +\epsilon^{2}\, \big\{G(a_{1};z)\, [\,G(a_{3};w)-G(a_{7};x)+G(a_{20};x)\,] \nonumber \\
&-\,G(a_{3};w)\, [\,G(a_{1};x)+G(a_{8};y)\,]-G(a_{20};x)\, [\,G(a_{3};y)-G(a_{3};w)\,]-\,G(a_{1},a_{1};x) \nonumber \\
&+G(a_{1},a_{1};z)-G(a_{1},a_{3};w)+G(a_{3},a_{3};w)+G(a_{7},a_{1};x)+G(a_{8},a_{3};y) \nonumber \\
&\big.-G(a_{20},a_{1};x)\big\}\,, \nonumber \\
g_{43} ={}& \epsilon\, [\,G(a_{1};y)-G(a_{1};z)\,] -\epsilon^{2}\, \big\{G(a_{1};y)\, [\,G(a_{3};w)+G(a_{21};x)\,]-G(a_{1};z)\, [\,G(a_{3};w) \nonumber \\
&-G(a_{7};y)+G(a_{12};y)+G(a_{21};x)\,]+G(a_{3};w)\, [\,G(a_{8};x)-G(a_{12};y)-G(a_{21};x)\,] \nonumber \\
&+G(a_{1},a_{1};y)-G(a_{1},a_{1};z)+G(a_{1},a_{3};w)-G(a_{3},a_{3};w)-G(a_{7},a_{1};y) \nonumber \\
&\big.-G(a_{8},a_{3};x)+G(a_{12},a_{1};y)+G(a_{21},a_{3};x)\big\}\,, \nonumber \\
g_{44} ={}& \epsilon^{2}\, \big\{G(a_{1};z)\, [\,G(a_{3};w)-G(a_{7};x)+G(a_{20};x)\,]-G(a_{3};w)\, [\,G(a_{1};x)+G(a_{8};y)\,] \nonumber \\
&-G(a_{20};x)\, [\,G(a_{3};y)-G(a_{3};w)\,]-G(a_{1},a_{3};y)+G(a_{3},a_{3};w)+G(a_{7},a_{1};x) \nonumber \\
&\big.+G(a_{8},a_{3};y)-G(a_{20},a_{1};x)\big\}\,, \nonumber \\
g_{45} ={}& \epsilon^{2}\, \big\{G(a_{1};z)\, [\,G(a_{3};w)-G(a_{7};y)+G(a_{12};y)+G(a_{21};x)\,]-G(a_{1};y)\, [\,G(a_{3};w) \nonumber \\
&+G(a_{21};x)\,]-G(a_{3};w)\, [\,G(a_{8};x)-G(a_{12};y)-G(a_{21};x)\,]-G(a_{1},a_{3};x) \nonumber \\
&\big.+G(a_{3},a_{3};w)+G(a_{7},a_{1};y)+G(a_{8},a_{3};x)-G(a_{12},a_{1};y)-G(a_{21},a_{3};x)\big\}\,, \nonumber \\
g_{46} ={}& 0\,, \nonumber \\
g_{47} ={}& \epsilon^{2}\, \big\{G(a_{3};\mathbbm{w})\, [\,G(a_{9};y)-G(a_{18};y)\,]+G(a_{1},a_{3};y)+2 G(a_{3},a_{3};\mathbbm{w})-G(a_{9},a_{3};y) \nonumber \\
&\big.-G(a_{18},a_{3};y)\big\}\,, \nonumber \\
g_{48} ={}& \epsilon^{2}\, \big\{G(a_{3};\mathbbm{w})\, [\,G(a_{9};x)-G(a_{18};x)\,]+G(a_{1},a_{3};x)+2 G(a_{3},a_{3};\mathbbm{w})-G(a_{9},a_{3};x) \nonumber \\
&\big.-G(a_{18},a_{3};x)\big\}\,, \nonumber \\
g_{49} ={}& -\epsilon^{2}\, \big\{[\,G(a_{13};\mathbbm{x})-G(a_{25};\mathbbm{x})\,] [\,G(a_{2};\mathbbm{z})-G(a_{14};\mathbbm{z})+G(a_{15};\mathbbm{z})-G(a_{16};\mathbbm{z})\,] \nonumber \\
&-G(a_{3};\mathbbm{w})\, [\,G(a_{14};\mathbbm{x})+G(a_{14};\mathbbm{z})-G(a_{16};\mathbbm{x})-G(a_{16};\mathbbm{z})\,]+G(a_{3};y)\, [\,2 G(a_{13};\mathbbm{x}) \nonumber \\
&-G(a_{14};\mathbbm{x})+G(a_{14};\mathbbm{z})-2 G(a_{15};\mathbbm{z})-G(a_{16};\mathbbm{x})+G(a_{16};\mathbbm{z})\,]-G(a_{1},a_{13};\mathbbm{x}) \nonumber \\
&-G(a_{1},a_{15};\mathbbm{z})+G(a_{1},a_{25};\mathbbm{x})+G(a_{1},a_{26};\mathbbm{z})+G(a_{2},a_{15};\mathbbm{z})-G(a_{2},a_{27};\mathbbm{z}) \nonumber \\
&-2 G(a_{3},a_{3};y)+2 G(a_{3},a_{3};\mathbbm{w})-G(a_{13},a_{13};\mathbbm{x})+G(a_{13},a_{25};\mathbbm{x})+G(a_{14},a_{13};\mathbbm{x}) \nonumber \\
&-G(a_{14},a_{25};\mathbbm{x})-G(a_{14},a_{26};\mathbbm{z})+G(a_{14},a_{27};\mathbbm{z})+G(a_{15},a_{26};\mathbbm{z})-G(a_{15},a_{27};\mathbbm{z}) \nonumber \\
&\big.+G(a_{16},a_{13};\mathbbm{x})-G(a_{16},a_{25};\mathbbm{x})-\,G(a_{16},a_{26};\mathbbm{z})+G(a_{16},a_{27};\mathbbm{z})\big\}\,, \nonumber \\
g_{50} ={}& 0\,, \nonumber \\
g_{51} ={}& 0\,, \nonumber \\
g_{52} ={}& 8 \epsilon\, G(a_{3};y) +2 \epsilon^{2}\, \big\{2 G(a_{1};z)\, [\,G(a_{10};x)-G(a_{10};y)\,]-G(a_{3};\mathbbm{w})\, [\,G(a_{9};x) \nonumber \\
&-5 G(a_{9};y)-G(a_{18};x)+5 G(a_{18};y)\,]-2 G(a_{1},a_{3};x)+G(a_{9},a_{3};x) \nonumber \\
&-2 G(a_{10},a_{3};x)+G(a_{18},a_{3};x)+10 G(a_{1},a_{3};y)-16 G(a_{3},a_{3};y)-5 G(a_{9},a_{3};y) \nonumber \\
&\big.+2 G(a_{10},a_{3};y)-5 G(a_{18},a_{3};y)+8 G(a_{3},a_{3};\mathbbm{w})\big\}\,, \nonumber \\
g_{53} ={}& 0\,, \nonumber \\
g_{54} ={}& 0\,, \nonumber \\
g_{55} ={}& \epsilon^{2}\, \big\{G(a_{3};\mathbbm{w})\, [\,2 G(a_{1};x)-2 G(a_{1};z)+G(a_{9};y)+G(a_{18};y)\,]+3 G(a_{3},a_{3};\mathbbm{w}) \nonumber \\
&\big.-3 G(a_{3},a_{5};\mathbbm{w})-3 G(a_{3},a_{6};\mathbbm{w})-G(a_{9},a_{3};y)+G(a_{18},a_{3};y)\big\}\,, \nonumber \\
g_{56} ={}& \epsilon^{2}\, \big\{G(a_{3};\mathbbm{w})\, [\,2 G(a_{1};y)-2 G(a_{1};z)+G(a_{9};x)+G(a_{18};x)\,]+3 G(a_{3},a_{3};\mathbbm{w}) \nonumber \\
&\big.-3 G(a_{3},a_{5};\mathbbm{w})-3 G(a_{3},a_{6};\mathbbm{w})-G(a_{9},a_{3};x)+G(a_{18},a_{3};x)\big\}\,, \nonumber \\
g_{57} ={}& -\epsilon^{2}\, \big\{2\, [\,G(a_{1};x)-G(a_{1};z)+G(a_{19};y)-G(a_{23};x)\,]\, [\,G(a_{3};\mathbbm{w})-G(a_{5};\mathbbm{w}) \nonumber \\
&-G(a_{6};\mathbbm{w})\,]-2 G(a_{1};z)\, [\,G(a_{7};x)-G(a_{23};x)\,]+G(a_{3};\mathbbm{w})\, [\,G(a_{9};y)-G(a_{18};y)\,] \nonumber \\
&-2 G(a_{3};y)G(a_{23};x)+2 G(a_{1},a_{3};\mathbbm{w})-2 G(a_{1},a_{5};\mathbbm{w})-2 G(a_{1},a_{6};\mathbbm{w})+3 G(a_{3},a_{3};\mathbbm{w}) \nonumber \\
&-G(a_{3},a_{5};\mathbbm{w})-G(a_{3},a_{6};\mathbbm{w})-2 G(a_{5},a_{3};\mathbbm{w})+2 G(a_{5},a_{5};\mathbbm{w})+2 G(a_{5},a_{6};\mathbbm{w}) \nonumber \\
&-2 G(a_{6},a_{3};\mathbbm{w})+2 G(a_{6},a_{5};\mathbbm{w})+2 G(a_{6},a_{6};\mathbbm{w})+2 G(a_{7},a_{1};x)-G(a_{9},a_{3};y) \nonumber \\
&\big.-G(a_{18},a_{3};y)+2 G(a_{19},a_{3};y)-2 G(a_{23},a_{1};x)\big\}\,, \nonumber \\
g_{58} ={}& -\epsilon^{2}\, \big\{2\, [\,G(a_{1};y)-G(a_{1};z)+G(a_{19};x)-G(a_{22};y)-G(a_{24};x)\,]\, [\,G(a_{3};\mathbbm{w})-G(a_{5};\mathbbm{w}) \nonumber \\
&-G(a_{6};\mathbbm{w})\,]-2 G(a_{1};z)\, [\,G(a_{7};y)-G(a_{22};y)-G(a_{24};x)\,]+G(a_{3};\mathbbm{w})\, [\,G(a_{9};x) \nonumber \\
&-G(a_{18};x)\,]-2 G(a_{1};y)\, G(a_{24};x)+2 G(a_{1},a_{3};\mathbbm{w})-2 G(a_{1},a_{5};\mathbbm{w})-2 G(a_{1},a_{6};\mathbbm{w}) \nonumber \\
&+3 G(a_{3},a_{3};\mathbbm{w})-G(a_{3},a_{5};\mathbbm{w})-G(a_{3},a_{6};\mathbbm{w})-2 G(a_{5},a_{3};\mathbbm{w})+2 G(a_{5},a_{5};\mathbbm{w}) \nonumber \\
&+2 G(a_{5},a_{6};\mathbbm{w})-2 G(a_{6},a_{3};\mathbbm{w})+2 G(a_{6},a_{5};\mathbbm{w})+2 G(a_{6},a_{6};\mathbbm{w})+2 G(a_{7},a_{1};y) \nonumber \\
&\big.-G(a_{9},a_{3};x)-G(a_{18},a_{3};x)+2 G(a_{19},a_{3};x)-2 G(a_{22},a_{1};y)-2 G(a_{24},a_{3};x)\big\}\,, \nonumber \\
g_{59} ={}& 0\,,
\end{align}
where the weights of the GPLs are
\begin{align}
a_{1} &= 0\,, & a_{2} &= 1\,, & a_{3} &= -\,1\,, \nonumber \\
a_{4} &= -\,2\,, & a_{5} &= (-1)^{2/3}\,, & a_{6} &= -\,(-1)^{1/3}\,, \nonumber \\
a_{7} &= z\,, & a_{8} &= w\,, & a_{9} &= \mathbbm{w}\,, \nonumber \\
a_{10} &= z-1\,, & a_{11} &= 1+w\,, & a_{12} &= z\, (1+w)\,, \nonumber \\
a_{13} &= -\,\frac{1}{y}\,, & a_{14} &= -\,\frac{1}{\mathbbm{w}}\,, & a_{15} &= 1+\frac{1}{y}\,, \nonumber \\
a_{16} &= 1+\frac{1}{\mathbbm{w}}\,, & a_{17} &= \frac{z w}{y}\,, & a_{18} &= -\,\frac{\mathbbm{w}}{1+\mathbbm{w}}\,, \\
a_{19} &= \frac{\mathbbm{w}^{2}}{1+\mathbbm{w}}\,, & a_{20} &= \frac{z\, (1+w)}{1+y}\,, & a_{21} &= \frac{z\,(1+w)-y}{y}\,, \nonumber \\
a_{22} &= \frac{z\, (1+\mathbbm{w}+\mathbbm{w}^{2})}{1+\mathbbm{w}}\,, & a_{23} &= \frac{z \left(1+\mathbbm{w}+\mathbbm{w}^{2}\right)}{(1+y)\, (1+\mathbbm{w})}\,, & a_{24} &= \frac{z \mathbbm{w}^{2}+(z-y)\, (1+\mathbbm{w})}{y\, (1+\mathbbm{w})}\,, \nonumber \\
a_{25} &= \frac{-(1+y)\, (1+\mathbbm{w})}{y\, (1+\mathbbm{w})+\mathbbm{w}^{2}\,(1+y)}\,, & a_{26} &= \frac{(1+\mathbbm{w})\, (1+y \mathbbm{x})}{y\, (1+\mathbbm{w})-\mathbbm{w}^{2}\, (1-\mathbbm{x})}\,, \nonumber \\
a_{27} &\rlap{$\displaystyle {} = \frac{(1+y)\,(1+\mathbbm{w})+\mathbbm{x}\, [\,y\,(1+\mathbbm{w})+\mathbbm{w}^{2}\,(1+y)\,]}{y\, (1+\mathbbm{w})+\mathbbm{x} \mathbbm{w}^{2}\, (1+y)}\,.$} & & \nonumber
\end{align}

\bibliographystyle{apsrev4-2}
\bibliography{refs}
\end{document}